# Enemy At the Gateways: A Game Theoretic Approach to Proxy Distribution

Milad Nasr
University of Massachusetts Amherst
milad@cs.umass.edu

Sadegh Farhang
Pennsylvania State University
farhang@ist.psu.edu

Amir Houmansadr
University of Massachusetts Amherst
amir@cs.umass.edu

Jens Grossklags
Technical University of Munich
jens.grossklags@in.tum.de

***Abstract*—A core technique used by popular proxy-based circumvention systems like Tor, Psiphon, and Lantern is to secretly share the IP addresses of circumvention proxies with the censored clients for them to be able to use such systems. For instance, such secretly shared proxies are known as *bridges* in Tor. However, a key challenge to this mechanism is the *insider attack* problem: censoring agents can impersonate as benign censored clients in order to obtain (and then block) such secretly shared circumvention proxies.**

**In this paper, we perform a fundamental study on the problem of insider attack on proxy-based circumvention systems. We model the proxy distribution problem using game theory, based on which we derive the optimal strategies of the parties involved, i.e., the censors and circumvention system operators. That is, we derive the *optimal* proxy distribution mechanism of a circumvention system like Tor, against the censorship adversary who also takes his optimal censorship strategies. This is unlike previous works that design ad hoc mechanisms for proxy distribution, against non-optimal censors.**

**We perform extensive simulations to evaluate our optimal proxy assignment algorithm under various adversarial and network settings. Comparing with the state-of-the-art prior work, we show that our optimal proxy assignment algorithm has superior performance, i.e., better resistance to censorship even against the strongest censorship adversary who takes her optimal actions. We conclude with lessons and recommendation for the design of proxy-based circumvention systems.**

## I. Introduction

Internet censorship is a global threat to the freedom of speech, ideas, and information. An increasing number of repressive regimes and totalitarian governments censor their citizens' access to the Internet [31], [17], [21], [6], [9], [4] using techniques such as IP address filtering, DNS interference, and deep-packet inspection [20], [34]. To help censored users get around censorship a number of tools and techniques have being designed and deployed by practitioners and academics, which are broadly known as *censorship circumvention tools* [28], [8], [11], [3], [19], [26], [27]. Such tools range from classic VPNs and one-hop HTTP proxies to more advanced techniques such as domain fronting [11], [14] and decoy routing [16], [38], [18].

A **core technique** used by *most* of the widely-deployed circumvention tools is to *share some secret information* with the censored users for them to be able to use the circumvention system. A censored client can only use the circumvention system with knowledge of that secret information. That is, the secret information is the censored users' gateway to the free Internet. For instance, in proxy-based circumvention systems like Tor, Lantern, and Psiphon this shared information is the IP address of some secret proxy server (e.g., a Tor bridge [7] IP address), which enables a censored user to connect to the circumvention system as long as the secret IP address of that proxy is kept undisclosed to the censors (and therefore, non-blocked).[1] We refer to this problem as the *proxy assignment problem*.

While some circumvention systems have started to partly deploy *domain fronting* [11] to resist blocking of their circumvention proxies, domain fronting is —prohibitively expensive—to be deployed at large scale; therefore, the proxy distribution problem as stated above remains a major challenge to proxy-based circumvention systems.

Unfortunately, this widely-deployed approach (i.e., granting access based on secret information) is prone to a fundamental issue, which we call the *insider attack*. Major circumvention systems like Tor, Lantern, etc., are designed to serve the masses, i.e., they are open to anyone who claims to be censored. Therefore, censoring agents can impersonate censored users and join the system in order to learn the secret information (e.g., Tor bridge IP addresses), and consequently block the circumvention system.

To limit the damage from the insider attack, a circumvention systems should limit the secret information disclosed to the censoring agents (e.g., only disclose a small fraction of Tor bridges to the censors). However, distinguishing censoring agents from genuine censored users is extremely challenging to circumvention system operators, as they will connect from the same geographic regions, and are using the same circumvention software. The main technique deployed by circumvention systems to limit the damage by the insider attack is constraining access to the secret information for each requesting client. For instance, Tor restricts the number of bridge IPs shared with each censored client to three at each time. However, this mechanism is known to be *ineffective*. For instance, Chinese censors were able to enumerate *all* Tor bridges within a month [37]. Further, limiting the access to such information (e.g., reducing the number of Tor bridges shared with clients) is likely to impact the usability of the circumvention system for genuine censored clients, e.g., a genuine censored user may soon run out of working Tor bridge IPs.

[1]Note that some of these circumvention systems have started to partly deploy *domain fronting* [11] to resist blocking of their circumvention proxies, however, domain fronting is —prohibitively expensive—to be deployed at large scale; therefore, the proxy distribution problem as stated above remains a major challenge to proxy-based circumvention systems.

**Our contributions.** In this paper, we perform a fundamental study on the problem of the insider attack on circumvention systems. Without loss of generality and for simplicity, we present our solution for the proxy assignment problem, however it can be trivially expanded to the generic problem of sharing secret information in circumvention systems. Specifically, we investigate the problem of circumvention proxy distribution by finding the optimal strategies of the parties involved, and deriving the Nash equilibrium. We do so by modeling the proxy assignment problem with a game theoretic model. Therefore, a significant contribution of our work is deriving the optimal strategies of the censors and circumvention operators, as opposed to designing ad hoc proxy assignment mechanisms. Additionally, compared to prior studies [36], [23], [22] we consider a more realistic threat model by incorporating practical issues into our model, such as the geographic locations of the clients and censors, the past actions of clients and proxies, etc.

We use game theory to find the optimal strategies of a circumvention system operator in assigning proxies to the clients in order to optimize the effectiveness of the circumvention system (e.g., maximize the number of genuine censored users who can obtain non-blocked Tor bridge IPs), while the censors also take their best actions to maximize the censorship damage. We model the proxy distribution problem using a classic matching game called the *college admissions game* [12] whose goal it is to admit students into colleges based on the rankings provided by the students as well as colleges. We build a *proxy assignment game* by making an analogy between circumvention clients and students, as well as between proxies and colleges. We define various metrics based on real-world constraints of circumvention systems to enable the clients and the proxies rank each other in the proxy assignment game. Based on our game, we derive the optimal algorithm for proxy distribution as well as the optimal censorship strategy.

We perform extensive simulations to evaluate our optimal proxy assignment algorithm under various adversarial and network settings. Comparing with the state-of-the-art prior work, we show that our optimal proxy assignment algorithm has superior performance, i.e., better resistance to censorship even against the strongest censorship adversary who takes her optimal actions. We conclude with the lessons learned for the design of proxy assignment systems.

## II. Problem Statement and Threat Model

### A. The Insider Attack Problem

Many circumvention systems work by sharing some secret information with their clients, where the access to such secret information is key in being able to use those circumvention systems. For many systems, the shared information are the IP addresses of the proxy servers used for circumvention, e.g., Tor bridges [7]. While our analysis is generic to any circumvention system relying to some degree on secret information, in the rest of this paper we consider the shared secret information to be the IP addresses of circumvention *proxies* for simplicity purposes.

Figure 1 illustrates the main setting of proxy distribution in a circumvention system. The problem consists of the following entities:

- **Distributor:** A circumvention system entity which is in charge of distributing the secret information about scarce resources (e.g., proxy IP addresses) among clients. In the case of Tor, this entity is Tor's bridge distribution service.
- **Censored Clients:** The benign censored clients who are genuinely interested in using a circumvention system to sidestep censorship mechanisms. These users will ask the distributor entity for proxy information.
- **Censoring Agents:** The rogue clients controlled by the censorship authorities to impersonate real censored clients and obtain the secret proxy information from the circumvention distributor entity.
- **Censor:** The central censorship authority who collects and combines the information obtained by its censoring agents. The collected information (e.g., Tor bridge information) can be used by the censor to block the circumvention system.

In this setting, the distributor entity has the objective to disclose the lowest possible number of proxies to the censor. In contrary, the censor entity aims at identifying as many proxies as possible.

### B. Bridge Distribution in Tor

The Tor project offers three mechanisms to the clients to obtain Tor bridge IP addresses[2]: (1) a user's Tor client software (*Vidalia*) can directly obtain bridge IPs from Tor servers, (2) a user can visit Tor's bridge distribution webpage[3] and obtain bridge IPs after solving a CAPTCHA, and (3) a user can send an email to `bridges@bridges.torproject.org` from Gmail, Yahoo!, or Riseup! to receive bridge IPs via an email response.

To protect bridge information from the censoring agents, Tor limits the number of proxy IPs returned to a requesting client to three. Tor identifies users based on their IP addresses and email addresses depending on the mechanism used to obtain bridges. However, such protection mechanisms are ineffective against a resourceful censor who can create large numbers of email accounts or connect from a diverse set of IP addresses. In fact, the Chinese censors were able to enumerate all Tor bridges over the course of a single month [37].

Note that as mentioned earlier, the proxy distribution problem is not only an issue for Tor, but for all major proxy-based circumvention systems like Psiphon [28], Ultrasurf [35], and Lantern [19]. While some circumvention systems have started to partly deploy *domain fronting* [11] to resist blocking of their circumvention proxies, domain fronting is —prohibitively expensive—to be deployed at large scale; therefore, the proxy distribution problem as stated above remains a major challenge to proxy-based circumvention systems. For instance, due to the high cost [25] of running Tor's domain fronting bridges (meek [24]), recent proposals suggest to use meek only for bootstrapping, not for proxying circumvention traffic.

[2]https://www.torproject.org/docs/bridges
[3]https://bridges.torproject.org/

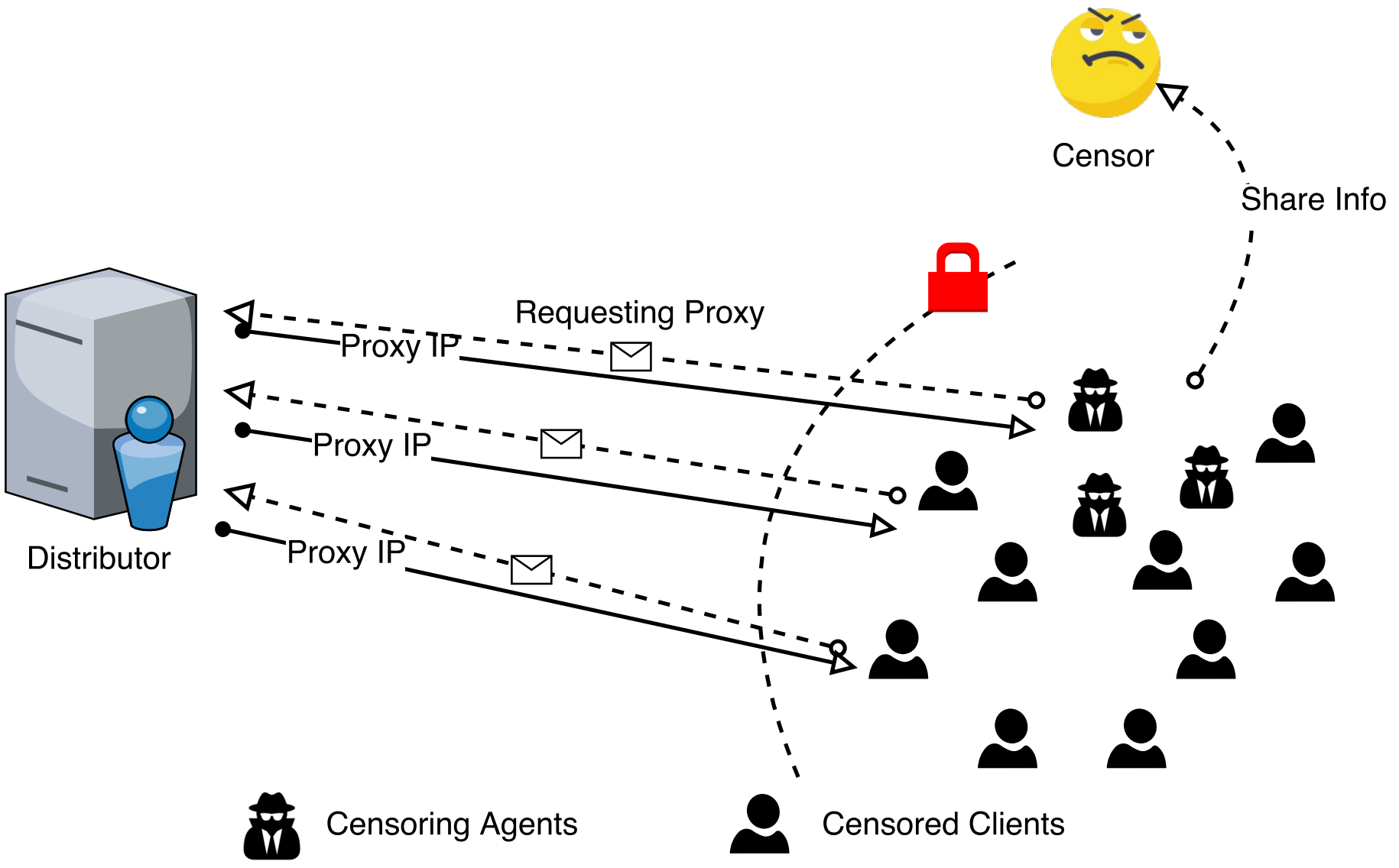

Fig. 1. Overview of proxy distribution in a circumvention system.

### C. Threat Model

We assume that the clients have no way of obtaining the proxy information other than contacting the distributor entity, e.g., a censored client will not be able to obtain a proxy IP by asking friends. Similarly, we assume that the censoring agents can only obtain proxy information by requesting the circumvention distributor. Therefore, we do not consider other adversarial means of discovering proxies such as probing suspicious IPs [37], zig-zag attacks [32], and other forms of active attacks [15], [13] in this work. Needless to say, future work can trivially include any of these techniques into our generic game theoretic model (i.e., by modifying our utility functions described in the following sections); we refrain from doing so in this work in order to avoid over-complicating the results.

Unlike the simplifying, unrealistic assumption made in prior works [36], [23], [22], we assume that the censoring agents are able to communicate among themselves and the central censorship authority decides on blocking certain bridges by aggregating information from all clients. Therefore, the censor can strategically allow an identified bridge to remain unblocked for a while, or block it instantly depending on the importance of the discovered bridge as well as the reputation of the censoring agent who discovered that bridge.

## III. Sketch of Our Approach

We start by summarizing previous attempts to study the proxy distribution problem in circumvention systems. We will then introduce the approach taken in this paper.

### A. Prior Studies

A comparatively small number of research papers have studied the problem of insider attacks in circumvention systems.

**Client puzzles.** As described earlier, one of the mechanisms to obtain Tor bridges is to visit a Tor website and to solve a CAPTCHA. However, CAPTCHAs are known to be trivially defeatable by resourceful adversaries who can hire human labor to solve them. Alternatively, several research papers [10], [5] have investigated the use of cryptographic puzzles to defeat enumeration by Sybil censors. Feamster et al. [10] provide evidence that this approach has limited impact against powerful censors.

**Client Reputation.** Some solutions like Proximax [23] leverage user relationships in online social networks like Facebook to distribute proxy information. A more recent mechanism is rBridge [36] that uses clients' reputation to distribute bridge information. In this approach, each user has an amount of credits that changes over time based on the uptime of the bridges she knows. To prevent the censors from enumerating all bridges, rBridge users can only obtain new bridge IPs by spending their (finite) credits. A user can also use her credits to invite new users to rBridge. Unfortunately, these studies make *unrealistic assumptions* to simplify their derivations. For instance, they make no distinction between different proxies, while in practice different proxies have different value to the censors and clients (e.g., censors prefer to block proxies who serve more clients). Also, they assume that the censoring agents act independently in blocking proxies.

**Theoretical bounds.** To the best of our knowledge, Mahdian [22] is the only prior work to investigate the problem of proxy distribution from a theoretical perspective. He finds a lower bound on the number of required proxies using techniques from information theory, however, also relies on a number of unrealistic assumptions to simplify the analysis. For instance, the author assumes that the number of censoring agents is known to the distributor, and the number of bridges and users are constant.

### B. Our Direction

We perform a fundamental study on the problem of proxy distribution in circumvention systems. We use game theory to find the optimal proxy distribution strategy of a circumvention system (as opposed to ad hoc mechanisms) in the presence of a censorship adversary who also uses her optimal strategy in discovery and blocking the proxies (as opposed to ad hoc censorship mechanisms). Our work takes a major step forward in the study of proxy distribution systems by including important real-world constraints in the model. We define utility functions for each of the players involved in the insider attack problem, including the censor, genuine censored clients, and proxy distributor. In particular, we include constraints like the location of clients, the clients' patterns of using proxies, and censors' and clients' preferences in obtaining new proxies. We include reputation metrics for clients into their utility functions, similar to previous work on reputation-based solutions. Our model is generic, and future work can extend it from a theoretical and practical perspective by considering additional key features tailored to different circumvention systems like the uptime of users, traffic volumes of users, or any other real-world constraints of specific circumvention systems.

**Paper's Outline:** The rest of this paper is organized as follows. We first introduce the college admissions game in Section IV, which is the theoretical approach we use to model and evaluate the insider attack problem. We define the utility functions of the entities involved in the proxy distribution problem in Section V, and describe our proxy assignment game in Section VI along with the optimal proxy distribution and censorship strategies. We describe our simulation setup in Section VII, and the simulation results are presented in Section VIII. We discuss our results and offer concluding remarks in Section IX.

## IV. Background: College Admissions Game

In this section, we provide an overview of the college admissions game framework by Gale and Shapley [12], which is the foundation for our game theoretic model. The college admissions game is also referred to as *deferred acceptance*[4] and *many-to-one matching game*. First, we describe the assignment criteria in the college admissions game. Second, we describe the college admissions algorithm and its characteristics.

### A. The Assignment Criteria

In the college admissions game, there are $n$ students and $m$ colleges. Each college has a capacity of $q_i$ students that can be admitted. Each student ranks the colleges based on her preferences. Note that each student omits those schools in her preference list that she would never choose under any circumstances. On the other side, each college ranks the students who have applied to that school based on the college's preferences. Similar to students, the college first eliminates the students who will not be admitted under any circumstances even if the college does not reach its capacity. The college admissions algorithm then derives assignments considering the preferences of both colleges and students; subject to the capacity of colleges. Note that there may exist different assignments of students to colleges. Here, we are focusing on assignments which are **stable**. We define *unstable* assignments in the following definition.

**Definition 1.** *An assignment of students to colleges is called* ***unstable*** *if there are two students* 1 *and* 2 *who are assigned to colleges* $a$ *and* $b$*, respectively, however, student* 2 *prefers college* $a$ *to* $b$ *and college* $a$ *prefers student* 2 *to* 1 *[12].*

Note that it is possible that different stable assignments exist. Then, the question is how to choose among different stable assignments; ideally the optimal one. The following gives the definition of an *optimal* assignment in the college admissions game.

**Definition 2.** *A stable assignment is optimal if every student is just as well off under it as under any other stable assignment [12].*

It is worth mentioning that the optimal stable algorithm is unique. In other words, there exists (if it exists at all) only one assignment that is optimal and stable. Moreover, in the above definition, the assignment is optimal from the students' point of view. In the following subsection, we provide an overview of an algorithm that preserves these two features.

### B. Deferred Acceptance Algorithm

In this subsection, we describe the *deferred acceptance* algorithm. As mentioned in Section IV-A, there exist some students that a college will not admit under any circumstances. Here, in the deferred acceptance algorithm, the assumption is made that these students are not allowed to apply for that college. By considering this assumption, the algorithm is as follows. First, all students apply to the college that is their first choice. A college with capacity of $q$ students places the $q$ students on the waiting list with the highest rank, or all the students who have applied if there are less than $q$ applicants. The college rejects the remainder of the students. The rejected students apply to their next choice. In a similar way, each college selects the top $q$ students from the students on its waiting list from a previous round and the new students who have applied to this college. The college chooses the top $q$ students and rejects the rest. This procedure terminates if each student is on a waiting list of a college or has been rejected by all colleges he had been permitted to apply. Finally, each college admits the students on its waiting list.

The following two theorems describe the characteristics of the deferred acceptance algorithm.

**Theorem 1.** *There always exists a stable assignment in the deferred acceptance algorithm [12].*

*Proof:* Note that the deferred acceptance algorithm gives an iterative procedure for assigning students to schools. Here, we claim that the deferred acceptance is stable. In doing so, let's suppose that student $a$ is not admitted by school $b_1$, but student $a$ prefers school $b_1$ to his admitted school $b_2$. This means that student $a$ has applied to school $b_1$ at some stage and student $a$ has been rejected in favor of some students that school $b_1$ prefers more. Then, it is obvious that school $b_1$ must prefer its admitted students compared to student $a$. Therefore, there is no instability in the deferred acceptance algorithm [12]. ■

[4] We use deferred acceptance and college admission interchangeably throughout the paper.

**Theorem 2.** *Every student is at least as well off under the assignment given by the deferred acceptance algorithm as he would be under any other stable assignment [12].*

*Proof:* We prove the optimality by induction. Let's consider student $a_1$ and school $b_1$ with $q$ capacity. If student $a_1$ is sent to a school under a stable assignment, we call that school a *possible* one for student $a_1$. Let's assume that at a stage in the deferred acceptance algorithm, no student has been rejected from a school which is possible for him. Let us assume that school $b_1$ has received $q$ applications from $q$ students, i.e., $a_2$,..., $a_{q+1}$, who are more qualified compared to $a_1$. As a result, school $b_1$ rejects student $a_1$. We need to show that school $b_1$ is impossible for student $a_1$. Note that each student $a_i$, where $i \in \{2, ..., q+1\}$, prefers school $b_1$ to all of the other schools except those schools that they have previously applied to and have been rejected. In a hypothetical assignment, let us assume that student $a_1$ is sent to school $b_1$ and all other students are sent to schools that are possible for them.

Note that there exists at least one student $a_i$ where $i \in \{2, ..., q+1\}$ who has to go to a school which is a less desirable school compared to $b_1$. It is straightforward to see that this is an unstable assignment due to the fact that both the school and the student are dissatisfied about this assignment. This hypothetical assignment is unstable and school $b_1$ is impossible for student $a_1$. We can conclude that the deferred acceptance algorithm only rejects students who cannot be admitted in any stable assignment. Hence, the resulting assignment is optimal [12]. ■

**Why we use this game in our model.** We use the college admissions game to establish an optimal mechanism to assign proxies to clients; we call our mechanism the *proxy assignment game*. On a conceptual level, we model the proxy assignment problem by using the college admission game as a foundation as follows: the clients (including the censoring agents) act as the students who are interested in learning the addresses of the proxies, and the proxies act as colleges as each of them has a finite (known) capacity for serving clients. Solving this game results in the assignment of proxies to clients.

It is important to note that the theory and concept of the college admissions game and the deferred acceptance algorithm is successfully applied in practice, and has stood the test of time [29]. Even predating the publication of the seminal paper by Gale and Shapley is the National Resident Matching Program used to place United States medical school students into residency training programs. Nowadays, over 40,000 applicants and 30,000 positions are part of the program on an annual basis. At the same time, over 60 matching programs for medical subspecialties follow similar processes [29]. More recently, the matching approach has also been used in the scenario of school choice in the New York and Boston public school systems [1]. Further, key insights of the general concept are used in diverse scenarios such as kidney exchanges [30]. Finally, the 2012 Nobel Memorial Prize in Economic Sciences was awarded to Roth and Shapley "for the theory of stable allocations and the practice of market design;" thereby validating the profound practical and theoretical impact of the work.

## V. Defining Utility Functions

### A. Main Model

In this paper, we propose a game-theoretic model for the proxy assignment problem. Towards this, we leverage the analogy between the college admissions game introduced above and our proxy assignment problem. Specifically, we formulate the proxy assignment problem as a college admissions game in which colleges, i.e., *proxies*, and students, i.e., *users*, rank each other based on their utilities (preferences). In doing so, we define utility functions for users and proxies as described in the following. We will then use the deferred acceptance (DA) algorithm, introduced in Section IV, to find the stable associations between users and proxies, i.e., the Nash equilibrium.

Similar to the college admissions game, we need to define metrics for both users and proxies to rank each other. In doing so, we define utility functions for these entities in our framework.

In our model, there are $n$ users in need of proxies to circumvent censorship who are represented by a set $\mathbb{A} = \{a_1, a_2, ..., a_n\}$. This set contains the IDs of all users in our system. Among these $n$ users, $m$ of them are censoring agents, denoted by $\mathbb{J} = \{j_1, j_2, ..., j_m\}$. All of these censoring agents are controlled by the censor. There are also $l$ proxies in the system denoted by a set $\mathbb{I} = \{i_1, i_2, ..., i_l\}$. Each user is provided $k$ proxies by the distributor at any request. We divide the time dimension into intervals called *stages* denoted by $t$ (the game starts at $t = 0$). All the actions by the players, such as asking for new proxies, blocking proxies, providing proxies, etc., are performed at the end of the stages, not during a stage.

In the following, we use the subscript $a$ to refer to a client, and the subscript $i$ to refer to a proxy. $t$ represents the time stage. Table I lists all the notations that we use throughout the paper.

**Possible Actions of Players:** In each stage of our game, a censored client can take one or both of the following actions: (1) use a proxy she already knows for browsing (for censoring agents to look like benign clients), and/or (2) issue a request for new proxies. This is shown in Algorithm 1. In addition to the above actions, a censoring agent also shares her obtained proxy addresses with the central censor entity, who will decide whether to block that proxy at that given stage. The strategy taken by a censoring agents is complex, which is discussed in Section VI-A. At each stage, each proxy decides if he wants to accept a requesting client or not. Note that the distributor entity plays the game on behalf of all of the proxies and tries to identify the optimal assignment between the proxies and the requesting clients (we assume that proxies trust the central distributor entity.)

TABLE I. NOTATIONS

| Variable | Definition |
|---|---|
| $t$ | stage of the game |
| $\mathbb{A}$ ($\mathbb{I}$) | Set of all users (proxies) |
| $\mathbb{J}(m)$ | Set (Number) of censoring agents |
| $n$ ($l$) | Number of users (proxies) |
| $C_i$ | Capacity of proxy $i$ |
| $c_i^t$ | Number of connected users to proxy $i$ up to stage $t$ |
| $B_i^t$ ($\mathbb{B}_i^t$) | Number of users who have the address of proxy $i$ |
| $b_i$ | Number of corrupt users having the address of proxy $i$ |
| $\gamma_{a,i}^t$ | indicator of user $a$ use of blocked proxy $i$ up to stage $t$ |
| $T_{a,i}^t$ | Time user $a$ uses proxy $i$ up to stage $t$ |
| $R_a^t$ | user $a$'s number of requests up to stage $t$ |
| $\delta_a^t$ | Number of blocked proxies that user $a$ has up to stage $t$ |
| $\tau_i^t$ | Total time of using a proxy $i$ up to stage $i$ |
| $d_{a,i}$ | Distance of user $a$ from proxy $i$ |
| $u_i^t(a)(u_a^t(i))$ | Proxy (client) utility |
| $\phi(a)$ | Utility of each censoring agent |
| $\Phi^t$ | Utility of the central censor |
| $\mu_b, \mu_s$ | Rate of new clients in birth interval (stable interval) |
| $\lambda_b, \lambda_s$ | Rate of new proxies in birth interval (stable interval) |
| $(\alpha_1, \alpha_2, \alpha_3, \alpha_4, \alpha_5)$ | Weighting factors in (4) |
| $(\beta_1, \beta_2, \beta_3)$ | Weighting factors in (3) |
| $(\omega_1, \omega_2)$ | Weighting factors in (9) |
| $\eta$ | Acceptance utility threshold for new proxy request |
| $\nu$ | Cost of losing a censoring agent |
| $\rho$ | Ratio of the censoring agents to the total number of clients |
| $\pi_1(i)$ | Number of censoring users connected to proxy (i) |
| $\pi_2(i)$ | Number of censoring users connected to proxy (i) with enough utility to request new one |
| $\pi_3(i)$ | Number of censoring users assigned to proxy (i) |
| $\pi_4(i)$ | Number of censoring users with utilities lower than acceptance threshold |
| $p$ | Probability of blocking a proxy for a conservative censor |
| $\overline{T}$ | Maximum time a user can use proxy for utility improvement |

**Algorithm 1** A benign censored client's strategy

```
if Can connect to previous proxy then
    return
else
    Request for new proxy
    Remove proxy from your pool
    while Know any proxy do
        if Can connect then
            return
        end if
        Remove proxy from your pool
    end while
end if
```

### B. Metrics to Distinguish Censors from Clients

We use the following metrics to distinguish censoring agents from genuine censored clients:

**Blocked proxy usage ($\gamma_{a,i}^t$):** While each client may know multiple proxy IP addresses, she will likely use only one of them at any point in time. Therefore, if one of her unused (idle) proxies is blocked by the censor, a benign client may not immediately request new proxies, while a censoring agent may continuously ask for new proxies. We define the metric $\gamma_{a,i}^t$ to indicate if a user $a$ who has been using the proxy $i$ up to time $t$ will request a new proxy:

$$\gamma_{a,i}^t = \begin{cases} 1 & \text{if user } a \text{ has used proxy } i \\ & \text{up to stage } t \text{ and request a new proxy,} \\ 0 & \text{Otherwise.} \end{cases} \quad (1)$$

**Proxy utilization ($T_{a,i'}$):** Typically, a censoring agent will not use a proxy, which she has obtained from the distribution, while a genuine censored client will likely utilize her obtained proxies to circumvent censorship. Therefore, the utilization of the obtained proxies can be taken into consideration as a factor to distinguish genuine clients and censoring agents. Of course, a censoring agent can also use the obtained proxies in order to look like genuine clients. This, however, will be costly to the censors. (Real-world censors are not known to be doing so.) We use the metric $T_{a,i}^t$ to represent the duration of time a user $a$ has used the proxy $i$ during the total time interval of $t$.

**Number of requests for new proxy addresses ($R_a^t$):** We use the metric $R_a^t$ to represent the number of requests that user $a$ has made up to stage $t$ for new proxies. A benign client will typically ask for new proxies only once all of her proxies are blocked, whereas a censoring agent is likely to request new proxies more frequently to expedite its proxy discovery.

**Number of blocked proxies that a user knows ($\delta_a^t$):** We use the metric $\delta_a^t$ to represent the number of blocked proxies that user $a$ has known up to stage $t$. The metric is expected to be larger for censoring agents than genuine clients.

**Client locations ($d_{a,i}$):** We use $d_{a,i}$ to indicate the distance of user $a$ from proxy $i$ (in practice, this distance is estimated based on IP addresses). We normalize $d_{a,i}$ to the range $[0, 1]$. As discussed later, we use this distance metric to optimize performance and censorship resistance in our utility functions.

### C. Metrics to Rank Proxies

We also define the following metrics to compare the importance of various proxies. Censors are interested in learning (and blocking) the more important proxies, so the distributor should be more protective of the more valuable proxies.

**Number of users who know a proxy ($B_i^t$):** The number of users having the address of proxy $i$ at stage $t$ is denoted by $B_i^t$.

**Number of users connected to a proxy ($c_i^t$):** This is the number of users connected to a proxy $i$ at stage $t$.

**Total time utilization of a proxy ($\tau_i^t$):** Another metric to quantify the importance of a proxy is the sum of time intervals it has been used by different users. We use the following metric:

$$\tau_i^t = \sum_{a \in \mathbb{B}_i^t} T_{a,i}^t, \quad (2)$$

where $\mathbb{B}_i^t$ is the set of user IDs that have the address of proxy $i$ up to stage $t$. A higher value of $\tau_i^t$ means the proxy is more important for circumvention.

### D. Utility Functions

In this section, we derive the utility functions of the users and proxies in the proxy assignment problem based on the metrics introduced above. Suppose that a client $a \in \mathbb{A}$ has requested proxy $i \in \mathbb{I}$ at stage $t$.

**Client's utility.** We define a client's utility as:

$$u_a^t(i) = \left(\beta_1 B_i^t + \beta_2 c_i^t + \beta_3 \tau_i^t\right)^{\left(\frac{1}{d_{a,i}}\right)}. \quad (3)$$

in which, the user's utility is weighted with the proxy importance factors introduced in Section V-C, i.e., $B_i^t$, $c_i^t$, and $\tau_i^t$. That is, the user's utility will be higher if he chooses (and is assigned) the more important proxies (since they are more reliable). In the above utility function, $\beta_1$, $\beta_2$, and $\beta_3$ are the scaling factors indicating the relative importance of each of the proxy metrics. We also use the (normalized) distance metric, $d_{a,i} \in [0,1]$, in the exponent of our utility function for two reasons. First, the distance metric helps clients to be assigned to proxies that are closer to them, therefore, improving the quality of connection (that is, clients prefer proxies with better performance). Second, the distance metric increases the chances of assigning the same proxies to the clients in the same neighborhood. Therefore, this improves resilience against censors who are running censoring agents within the same subnet. Note that putting the distance metric in the exponent, the assignment algorithm prioritizes the location of clients before considering the other metrics.

**Proxy's utility.** The proxy distributor has three objectives: (1) assigning as many censoring agents as possible to the *same* set of proxies, (2) assigning censored users to reliable (non-blocked) proxies, and (3) keeping the proxies alive as long as possible. To achieve these three objectives, we use our defined parameters in the previous subsection to define a utility function for each proxy. The utility of proxy $i \in \mathbb{I}$ at stage $t$ for user $a \in \mathbb{A}$ is as follow:

$$\begin{aligned} u_i^t(a) = \Bigg(&\alpha_1 \min\Big(\sum_{i' \in \mathbb{I}} T_{a,i'}, \overline{T}\Big) - \alpha_2 R_a^t \\ &-\alpha_3 \sum_{i' \in \mathbb{I}} (1 - \gamma_{a,i'}^t) - \alpha_4 \delta_a^t + \alpha_5\Bigg)^{\left(\frac{1}{d_{a,i}}\right)}. \end{aligned} \quad (4)$$

As can be seen, the defined utility function uses the client's metrics defined in Section V-B to rank clients based on their chances of being censoring agents versus benign censored clients $(\alpha_1, \alpha_2, \alpha_3, \alpha_4)$ are the scaling factors weighing the significance of different metrics whose choices will be discussed in Section VII-A6). We also use the distance metric in the exponent as discussed above for client's utility function. In our formula, we use the minimum for uptime metric to prevent a censoring agent from cheating by making a large utility value through using a single proxy for a long time.

We assume that a benign client will not ask for new proxies before using his previosuly-obtained proxies. Therefore, we use a very large value for $\alpha_3$ compared to the other scaling factors. $\alpha_5$ is the initial utility for new users. We will later discuss the values of the scaling factors $(\alpha)$ in more details.

## VI. Playing the Proxy Assignment Game

In this section, we show how users and proxies rank each other based on our defined utility functions in Section V. We then derive the optimal attacker model in the proxy assignment game, and present our optimal proxy assignment algorithm.

Note that the proxy assignment game is virtual, meaning that the central distributor plays on behalf of all of the proxies and clients. In particular, a client does not issue a request for specific proxy addresses, since that user does not know the identities of the proxies. Instead, a user only requests *some* new proxy addresses and the distributor who knows all of the proxies gives that user one or more new proxy address(es). Thus, the proxy admissions game takes place when a distributor receives a certain number of proxy requests at stage $t$. The distributor then plays the game on behalf of both users and proxies by calculating the utility of all proxies and all users requesting new proxies according to the utility functions of (4) and (3). As a final step, the distributor assigns proxies to the clients requesting new proxy addresses based on the deferred acceptance algorithm.

A preference relation $\succeq_i$ for a proxy (client) is defined over the set of all clients (proxies). This relation is a binary relation which is complete, reflexive, and transitive [12]. By using these preference relations, proxies and users can rank each other. A proxy $i \in \mathbb{I}$ will rank all users making requests at stage $t$. In doing so, for any two users $a, a' \in \mathbb{A}$ and $a \neq a'$, we define the following preferences for a proxy $i \in \mathbb{I}$:

$$a \succeq_i a' \;\Rightarrow\; u_i^t(a) \geq u_i^t(a'), \quad (5)$$

where $u_i^t(.)$ is given by (4). The user with the highest utility according to (4) is the most preferred user for proxy $i$.

Similarly, each user $a \in \mathbb{A}$ uses the following preference relation $\succeq_a$ to rank proxies $i, i' \in \mathbb{I}$ $(i \neq i')$:

$$i \succeq_a i' \;\Rightarrow\; u_a^t(i) \geq u_a^t(i'), \quad (6)$$

where $u_a^t(i)$ is given by (3).

Note that in a college admissions game, it is desirable to have strict preferences (denoted by $\succ$). Here, we assume that when a player is indifferent between two choices, that player ranks these two choices randomly, e.g., by tossing a coin. Furthermore, in a college admissions game, each college can have a threshold for accepting new students. Here, we set a global threshold for all proxies. The utility of each user should be at least more than $\eta$ to be able to request a new proxy. That is, the distributor computes (4) without taking into account the distance metric $d_{a,i}$ and compares it to the threshold $\eta$. If it is more than $\eta$, the distributor accepts the request and uses (4) to compute the stable assignment.

We use the deferred acceptance algorithm for proxy assignment for censorship circumvention since it provides a stable assignment. Based on Theorem 1, the deferred acceptance algorithm guarantees that in the resulting assignment at the end of each stage, there is no user, regardless of his type, who prefers another proxy where that proxy also prefers that user. Further, according to Theorem 1, the resulting assignment is optimal from the user's perspective. In other words, if there are some users that share their proxies with each other, they do not have any incentive to change their proxies with other. For censoring agents considering the fact they are controlled by the censor. They share their proxy addresses with each other, but none of them have incentive to use other censoring agents. Considering all users are rational, none of them have incentives to use the proxies that other users have. Therefore, there is no need for each proxy to save the identity of users to examine whether they use other users' proxies.

### A. Optimal Censorship Strategy

In this section, we derive the optimal attack strategy for the censor. Unlike previous work that assumes each of the censoring agents to act independently, we consider a more realistic, stronger threat model in which the central censorship authority decides the actions to be taken for all of the censoring agents in order to maximize the censorship damage. Note that unlike the distributor entity, the censor entity does not know the list of all proxies in the system. Consider $\mathbb{P}$ to be the set of proxies (out of all existing proxies) that are known to the censor (i.e., obtained by its censoring agents). The censor entity uses utility functions similar to those used by the distributor (which we will describe below) in order to decide its optimal strategy, i.e., one that maximizes the utility functions of its censoring agents. Particularly, the censor uses the following utility function to rank the proxies he knows ($\mathbb{P}$):

$$u^t(a) = \left( \alpha_1 \min(\sum_{i' \in \mathbb{I}} T_{a,i'}, \overline{T}) - \alpha_2 R_a^t \right. \\ \left. -\alpha_3 \sum_{i' \in \mathbb{I}} (1 - \gamma_{a,i'}^t) - \alpha_4 \delta_a^t + \alpha_5 \right) \tag{7}$$

This is similar to the utility function used by the distributor in (3). It excludes the distance metric $d_{a,i}$ as the censor does not care about the distance of the proxies to be blocked.

Note that losing a censoring agent is costly to the censor (we quantify it with $\nu$), so he tries to minimize such losses. The censor loses an agent if the agent's utility goes below the acceptance threshold ($\eta$). We, therefore, extend the utility equation as follow:

$$\phi(a) = \begin{cases} u^t(a) & u^t(a) \geq \eta \\ -\nu & o.w \end{cases} \tag{8}$$

The optimal attacker can likely infer the coefficients in the equations from the behavior of the distributor over time. Therefore, in our simulations we used the same values for the attacker and the distributor.

An optimal attacker aims at maximizing his total utility as well as block as many benign users as possible. In other words, in each stage of the assignment, the attacker wants to maximize the following metric:

$$\Phi^t = \omega_1 \sum_{a \in \mathbb{J}} \phi^{t+1}(a) - \omega_2 \sum_{i \in \mathbb{I}} c_i^{t+1} \tag{9}$$

Note that on the right hand side of this equation, the stage is equal to $t+1$ rather than $t$. Here, the optimal attacker wants to increase the total of his utility in the next stage in order to receive a better ranking from the distributor's point of view. Further, the optimal attacker aims to decrease the number of connected users to all proxies in the next step. Therefore, we see this factor with a negative sign in the above formula. Note that $\omega_1$ and $\omega_2$ are constants representing the relative importance of these two factors. One can tune the values of the scaling factors in (9) to match it to different kinds of real-world censors, i.e., to make the censor more or less aggressive.

Finally, the optimal attacker selects a set of proxies to block in such a way as to maximize (9). This is known to be a NP problem, however, since each user is connected to only one proxy at a time, the attacker can use the independence of proxies to break down the equation for each proxy. Therefore, we have $\Delta\Phi^t = \Phi^t - \Phi^{t-1} = \sum_{i \in \mathbb{P}} \Delta\Phi^t(i)$, where

$$\Delta\Phi^t(i) = \omega_2 c_i^t - \omega_1 \left( (\alpha_1 + \alpha_2)\pi_1(i) \right. \\ \left. -\alpha_1 \pi_2(i) + \alpha_4 \pi_3(i) + \nu \pi_4(i) \right). \tag{10}$$

$\pi_1$ indicates the number of the censoring agents connected to proxy $i$ who have enough utility to request new proxies and increase their utilities. $\pi_2$ shows the number of censoring agents who were connected to proxy $i$ and have enough utility to request a new one, but cannot earn utility from that proxy (i.e., they get the maximum utility from that proxy). $\pi_3$ is the number of censoring agents who have been assigned to proxy $i$, and $\pi_4$ is the number of agents whose utilities are below the acceptance threshold. As said before, a rational distributor will set $\alpha_3$ to a very high value and as a result, the censoring agents never request new proxies. Therefore, the attacker only considers requesting proxies to replace proxies that his agents have been connected to. This is the reason why we do not have $\alpha_3$ in this equation. In (10), the attacker can compute $\pi_1, \pi_2, \pi_3$ by considering $i$ to be a blocked proxy. By computing $\Delta\Phi^t(i)$ for each proxy, the censor will determine the proxies to get blocked. If $\Delta\Phi^t(i)$ is positive, the attacker blocks the $i$th proxy, otherwise he will leave it unblocked.

As each censoring agent likely knows multiple proxies, the censor also needs to decide for each censoring agent to which of these known proxies the agent should connect to. It is reasonable for the censor to maximize the number of unique proxies his agents are connected to in order to block a higher number of benign clients. To determine which proxies the censoring agents should be connected to, we model the problem as a matching problem [2]. Each censoring agent has a set of proxies and we want to assign each agent to one proxy in order to maximize the number of unique proxies. In particular, we model this using the maximum cardinality matching on a bipartite graph [2]. To solve this problem, the attacker can use the Hopcroft Karp algorithm [2]. If an agent has no unique proxy to connect to, that agent chooses one of its available proxies at random.

### B. Optimal Proxy Distribution Strategy

Algorithm 2 summarizes the *optimal proxy assignment mechanism* used by the distributor to assign proxies to the requesting clients. Recall that the distributor plays the assignment game on behalf of all users. Benign censored clients will only petition the distributor for new proxies once all of the proxies they know are blocked. At the end of each stage of the game, the distributor will collect all of the proxy requests and run the assignment game, described above, to map proxies to requesting clients.

**Algorithm 2** Optimal proxy assignment.

The new users are initially assigned to new $k$ proxies
Users request new proxies
For each user, the distributor builds a preference list based on (5)
For each proxy, the distributor builds a preference list based on (6) (each proxy rejects the users whose utilities are less than the threshold)
The distributor runs the deferred acceptance algorithm [12] to get the stable assignment
The distributor assigns new proxies to the users

## VII. Simulations Setup

To evaluate our proxy assignment algorithm, we implemented a proxy system simulator in Python. Apart from the censor and the distributor algorithms, it is essential to derive the various parameters of our model, including the number of users, the ratio of censoring agents to censored users, the number of proxies, as well as various scaling factors in the utility functions.

### A. Simulation Parameters

We use Tor, presumably the most popular proxy-based circumvention system, as an example of a proxy-based circumvention system in order to derive some of the parameters. Tor is primarily an anonymization system, but it has been extended to be used as a circumvention system by introducing Tor's non-public relays called *bridges*. Therefore, we only consider Tor's bridges (but not its public relays) as the "proxies" in our proxy assignment game. Needless to say, the parameters can be adjusted to other systems and threat models.

*1) System's lifespan:* Figure 2 presents the number of active Tor bridges per month over time (error bars show standard deviation during each month). The figure shows a monotonic increase in the number of bridges during the first few years of their inception, which has changed in recent years due to various social and political events like the post-Snowden effect. Suggested by this figure, we divide the lifespan of a circumvention system into two phases. The *birth interval* is the initial phase of the circumvention system's operation, i.e., until it reaches a stable rate of growth. The second phase is the *stable interval*, which starts right after the birth interval. In our simulator, we define each time unit as *one day* in real world. We set the birth interval to 365, therefore, representing 1 year in real world.

*2) Ecosystem:* Figure 3 shows the number of Tor bridge users per month over time, showing different rates of increase in different intervals. Inspired by this, we define the following parameters to model Tor's ecosystem:

- $\mu$: The rate of new clients per time unit. $\mu_b$ is the rate during the birth interval, and $\mu_s$ is the rate after the birth interval.

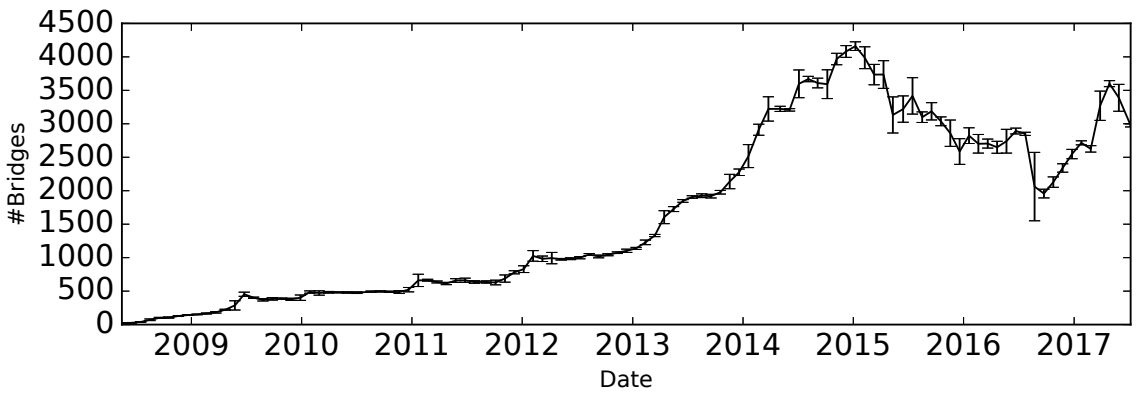

Fig. 2. Number of Tor bridges per month [33]

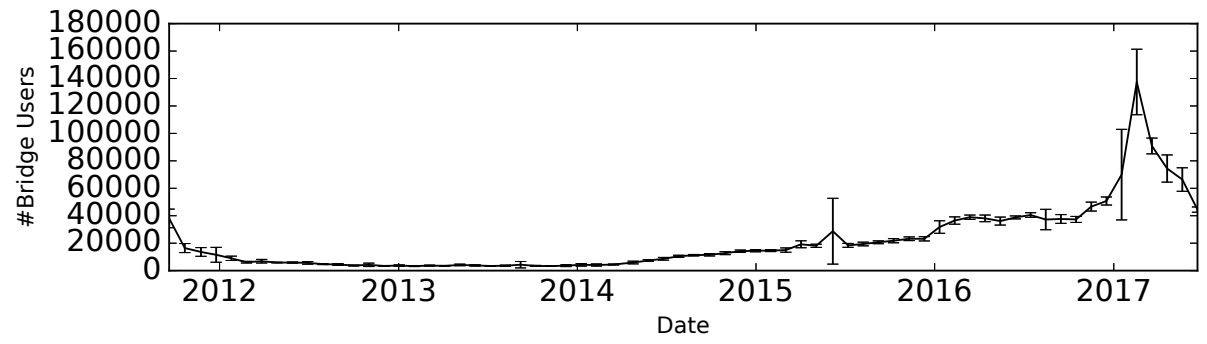

Fig. 3. Number of Tor bridge users per month [33]

- $\lambda$: The rate of new proxies per time unit. Similarly, $\lambda_b$ is the rate during the birth interval, and $\lambda_s$ is the rate after that.

We also define the parameter $\rho$ to denote the ratio of the censoring agents to the total number of clients. When adding a new user to the system, she will be a censoring agent with the probability $\rho$, otherwise a benign client.

To consider different settings, we define various ecosystems each with different values of the mentioned parameters. Table II lists the ecosystems used in our simulations.

*3) The distribution of censoring agents:* We consider two different types of censors. An *omnipresent censor* is a resourceful censor who is able to run censoring agents at various geographic locations (which therefore appear more like normal clients). On the other hand, a *circumscribed censor* is one who is running its censoring agents within a limited region, i.e., inside a single subnet.

*4) Users' locations:* In our simulations, we model the world as a $X \times X$ rectangular map with coordinates from $(-\frac{X}{2}, -\frac{X}{2})$ to $(\frac{X}{2}, \frac{X}{2})$. The censorship region covers a rectangular region from $(-y, -y)$ to $(y, y)$. Genuine censored users are uniformly distributed within the censored region, and the proxies are uniformly distributed outside of the censorship region. For the omnipresent censor, the censoring agents are distributed similar to benign users (i.e., uniformly). For the circumscribed censor, the censoring agents are distributed uniformly in a rectangular from $(-y_1, -y_1)$ to $(y_1, y_1)$, where $y_1 < y$. In our experiments, we set $X = 20000$, $y = 1000$, and $y_1 = 100$.

We use the Euclidean distance to determine the distance between the users and proxies. We normalize the distance metrics to the range $[0, 1]$.

*5) Proxy parameters:* Similar to Tor, the distributor returns 3 proxies to each new client. For existing clients, the distributor uses the game as discussed above to assign bridges. Also, without loss of generality, we set the capacity of each proxy to be 40 clients.

TABLE II. DIFFERENT CENSORSHIP ECOSYSTEMS

| World Name | $\mu_b$ | $\lambda_b$ | $\mu_s$ | $\lambda_s$ |
|---|---|---|---|---|
| Static | 25 | 5 | 0.1 | 0 |
| Slow | 25 | 5 | 5 | 0.2 |
| Alive | 25 | 5 | 10 | 0.5–7.5 |
| Popular | 25 | 5 | 20 | 0.5–10 |

TABLE III. VALUES FOR CONSTANTS

| Constants | Relative Values |
|---|---|
| $(\alpha_1, \alpha_2, \alpha_3, \alpha_4)$ | (Small, Very Small, Very Large, Large) |
| $(\beta_1, \beta_2, \beta_3, \eta)$ | (Small, Large, Large) |
| $(\eta, \alpha_5)$ | (Small, Large) |
| $(\omega_1, \omega_2, \nu)$ | (Small, Large, Very Large) |

*6) Scaling factors:* Table III shows the relative values of the scaling parameters of our utility functions ((3) and (4)), as used in our experiments. In (4), we use a small value for $\alpha_1$ to let benign users increase their utility by using their proxies over long time periods. The small value of $\alpha_1$ also prevents the censoring agents from increasing their utility of (4) by excessively using their proxies. We set $\alpha_2$ to a very small value as it is common for censored clients to request new proxies. We set $\alpha_3$ to a very large value since asking for a new proxy while having an unused proxy will be suspicious to the distributor. Finally, we set $\alpha_4$ to a large value to punish users who have many of their assigned proxies blocked. In our experiments, we use $(\alpha_1, \alpha_2, \alpha_3, \alpha_4) = (1, 1, 100, 5.0)$. We also set $\overline{T}$ to 100 ($\overline{T}$ denotes the maximum value).

In (3), we set $\beta_1$ to be smaller than $\beta_2$ since a proxy is more valuable if it has a large number of clients who are using it. Similarly, we set $\beta_3$ to be large to give more preference to the uptime of proxies. This also encourages the distributor to assign older proxies with higher priority. In our experiments, the values for these constants are $(1, 5, 5)$.

$\eta$ is the *acceptance threshold* of requests by proxies, and $\alpha_5$ is the initial utility of a client. Obviously, the acceptance threshold should be less than the initial utility; we use 0 and 10, respectively.

Finally, in (9), on the one side we have the sum of the utility points that a censoring agent loses by blocking proxy $i$, and on the other side the utility benefit a censor gets by blocking that proxy. We assume that blocking users is more important to the censors than keeping the censoring agents alive, therefore we set $\omega_2$ to be larger than $\omega_1$. We also use a large value for the cost of losing agents, $\nu$. We use 1, 100, and 500, respectively.

### B. Evaluation metrics

In our experiments, we use the following four metrics to evaluate the performance of proxy assignment in each setting.

- **Number of connected censored users:** This is the number of censored clients who know an unblocked proxy, and that proxy has unused capacity to serve the client.
- **Ratio of connected censored users to total number of censored users:** This shows the fraction of censored clients who know unblocked proxies and can connect to them.
- **Total capacity of the proxies:** This shows the total (used and unused) capacity of all unblocked proxies.
- **Wait time:** This metric shows how many rounds a censored client should wait to receive unblocked proxies.

### C. Censorship Strategies Evaluated

In addition to the *optimal* censorship strategy designed in the previous section, we evaluate two other censorship strategies in our evaluations that represent the mechanisms of previous studies. Therefore, we use the following censorship strategies in our evaluations:

- **Aggressive censor**: In this model, each censoring agent will immediately block a new proxy that she has learned.
- **Conservative censor**: In this model, each censoring agent keeps the proxies she has learned alive for a certain amount of time in order to increase her utility of the system. We use the same utility function used in our game (equation (7)) to model the utility of independent censoring agents. A censoring agent will block proxies after this time interval with probability $p$. If the censoring agents cannot increase their utility of (7) by waiting longer, she will simply block the proxies with probability 1.
- **Optimal censor**: This is the optimal game-theoretic censorship strategy derived in Section VI-A.

Recall that in previous bridge distribution mechanisms, particularly the state-of-the-art in rBridge [36], the censoring agents act *independently* in obtaining and blocking proxies, i.e., they do not communicate among themselves. Therefore, we use the aggressive and conservative censorship strategies to model the censors of prior work, and compare them to the optimal game-theoretic strategy derived in this paper. One can simply define other types of adversaries for independent censoring agents.

## VIII. SIMULATION RESULTS

We evaluate our proxy assignment game based on the setup described in Section VII. We use the metrics defined in Section VII-B to evaluate the performance of our assignment. We evaluate our game for the different censorship ecosystems (Table II), against different blocking strategies (Section VII-C), and for different distributions of censoring agents (omnipresent vs. circumscribed).

### A. Static World

As shown in Table II, the "static world" is the censorship ecosystem in which no new proxies are added to the system over time. As intuitively expected, our experiments show that the circumvention system is inefficient as the censors can eventually discover a large fraction of the proxies. Figure 4 shows the performance metrics (Section VII-B) of our proxy assignment mechanism in the static world ecosystem for an aggressive censor (for different fractions of censoring clients, $\rho$). As can be seen, even for the non-optimal aggressive censor (which is the least strategic censor), the circumvention system is not able to keep up, and (even the weakest) censor is able to block a large fraction of the proxies, therefore, the connected ratio metric does not increase over time with new clients joining the system (e.g., for $\rho = 0.1$). We conclude that independent of the censorship strategy, a circumvention system needs to add new proxies over time to be able to keep up with the censors.

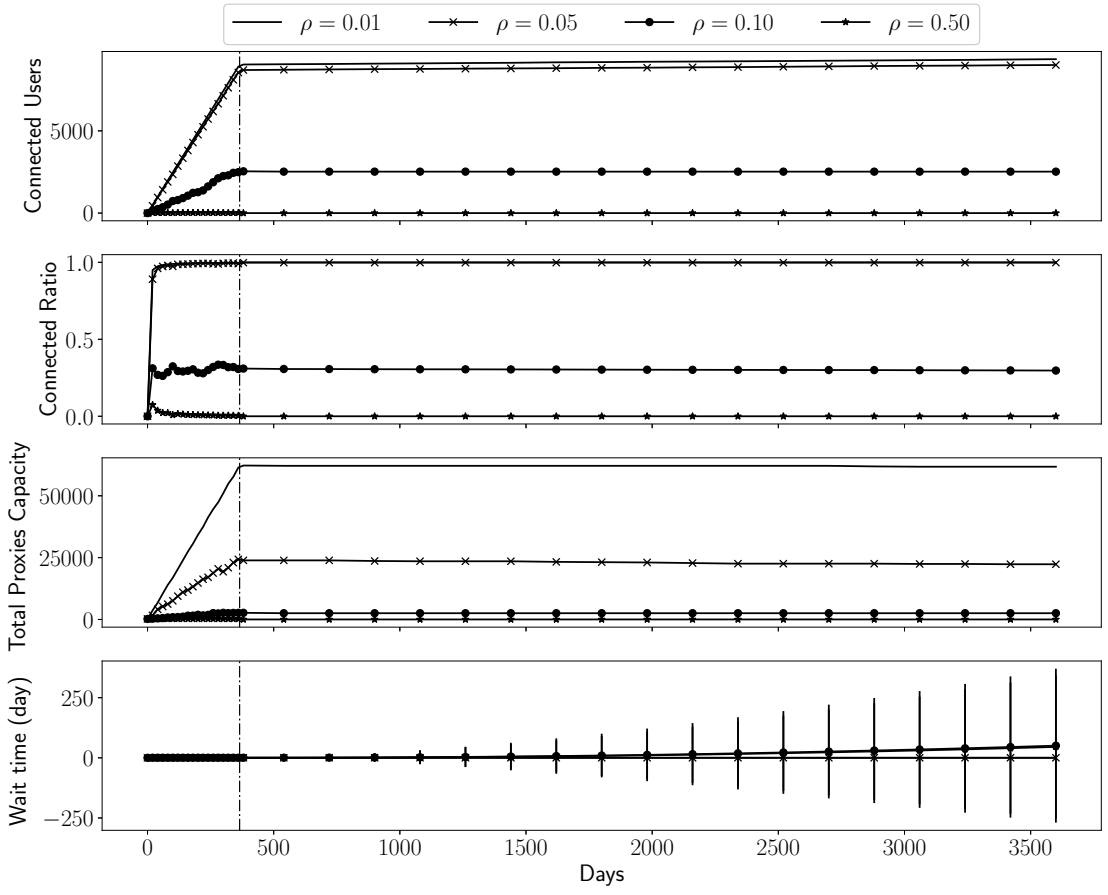

Fig. 4. Aggressive censor in a static world ecosystem.

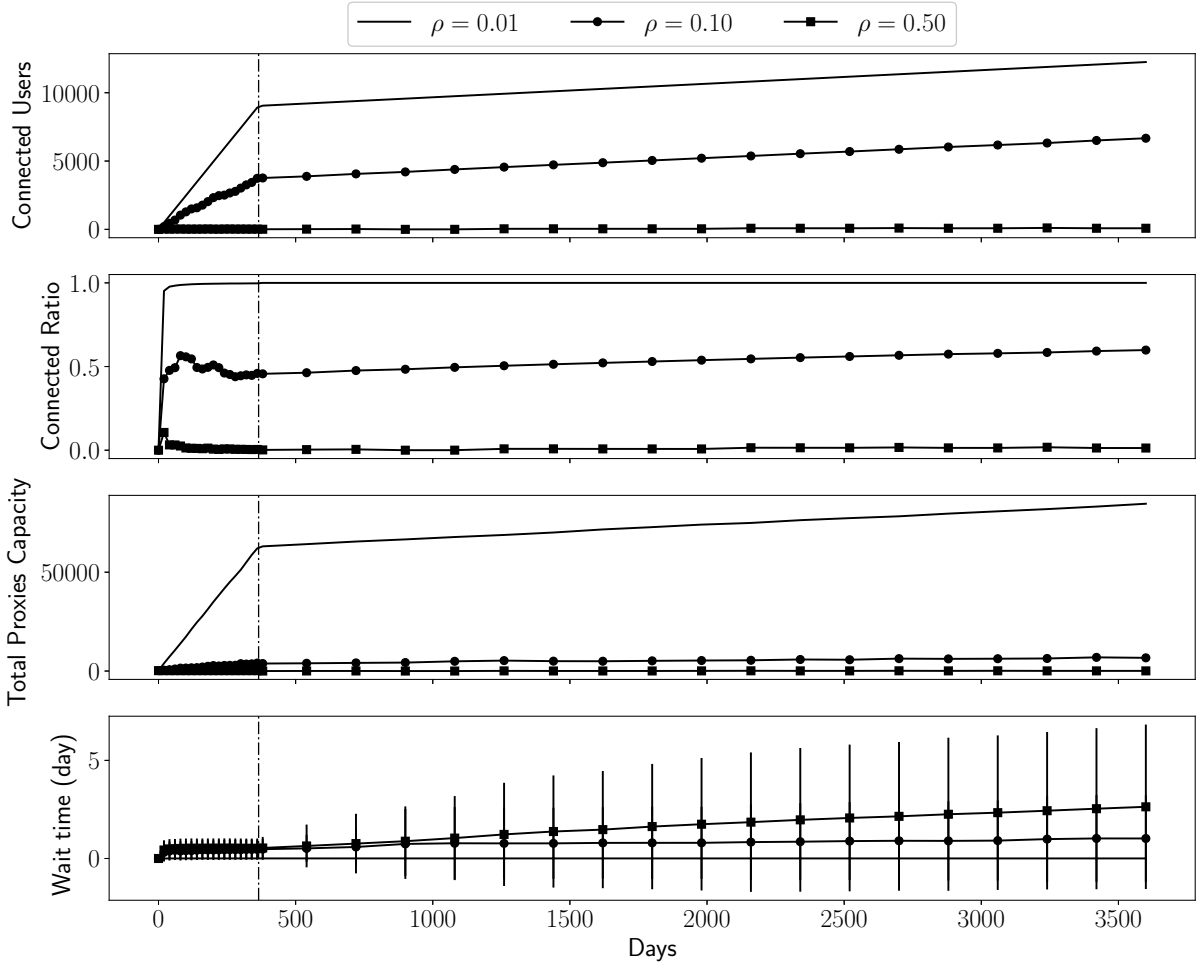

Fig. 5. Circumscribed censor with Aggressive blocking in Slow world

• **Lesson One:** Independent of the censorship strategy, a circumvention system needs to add new proxies over time to be able to keep up with the censors.

## *B. Different Distributions of Censoring Agents*

As mentioned earlier, we have two types of censoring agent distributions: *omnipresent* and *circumscribed*. Figures 5 and 6 compare the performance for these two distributions. As can be seen, the omnipresent type of censoring agents are more impactful since they can obtain a larger number of proxies due to their location diversity (which confirms (4)). However, note that it is costlier for the censors to distribute their censoring agents, therefore, the omnipresent censor represents a more resourceful censorship authority. In the following experiments, we will mainly use the omnipresent distribution as it represents a stronger censorship adversary.

• **Lesson Two:** Resourceful censors can increase their success by geographically distributing their censoring agents.

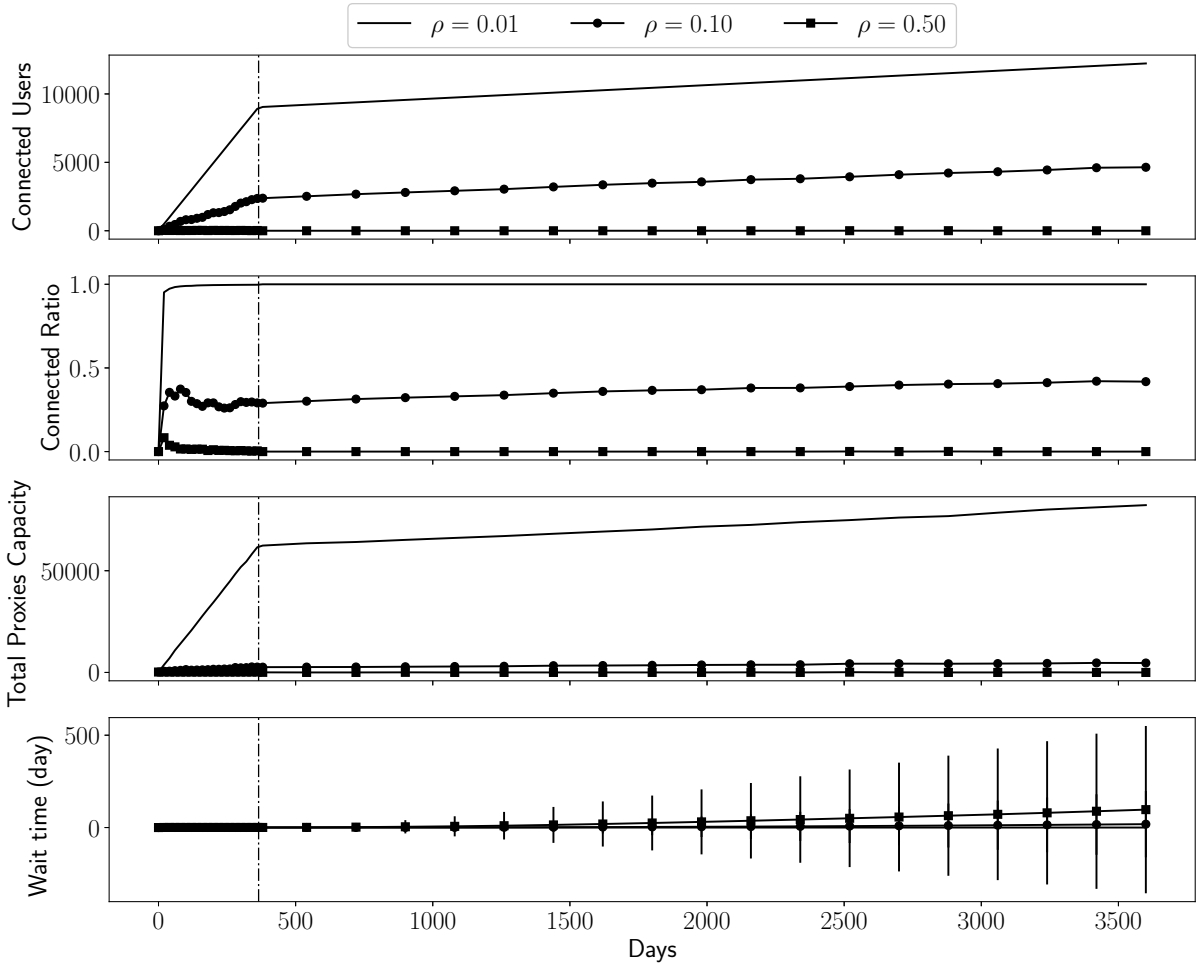

Fig. 6. Omnipresent censor with Aggressive blocking in Slow world

## *C. Comparing Censorship Strategies*

In Section VII-C, we introduced three strategies for the censors: optimal, which is the game-theoretic strategy derived in this work, and the two strategies of aggressive and conservative, which represent the mechanisms proposed by prior work. Figures 6, 7, and 8 compare the performance for aggressive, conservative, and optimal strategies, respectively. We observe that unlike previous works, our proxy assignment algorithm works better against a censor with a conservative blocking strategy than an aggressive one. This is because based on our utility functions, the conservative censoring agents do not gain any significant rewards due to longer wait times before blocking. Further, our utility functions give preference to older proxies by including a proxy's reliability (duration of operation) to rank proxies. Therefore, conservative censoring agents will likely get similar proxies that other censoring agents have obtained previously.

Also, comparing our optimal strategy to conservative and aggressive strategies, we find it to be significantly stronger, i.e., it blocks proxies more successfully than the ad hoc mechanisms of aggressive and conservative. We will use the optimal strategy in our following simulations as it is the strongest censorship strategy.

• **Lesson Three:** The censors can intensify their damage by applying strategic censorship mechanisms, as opposed to ad hoc ones.

## *D. Different Ecosystems*

We also compare the different ecosystems defined in Table II, each using different rates for adding new users and proxies.

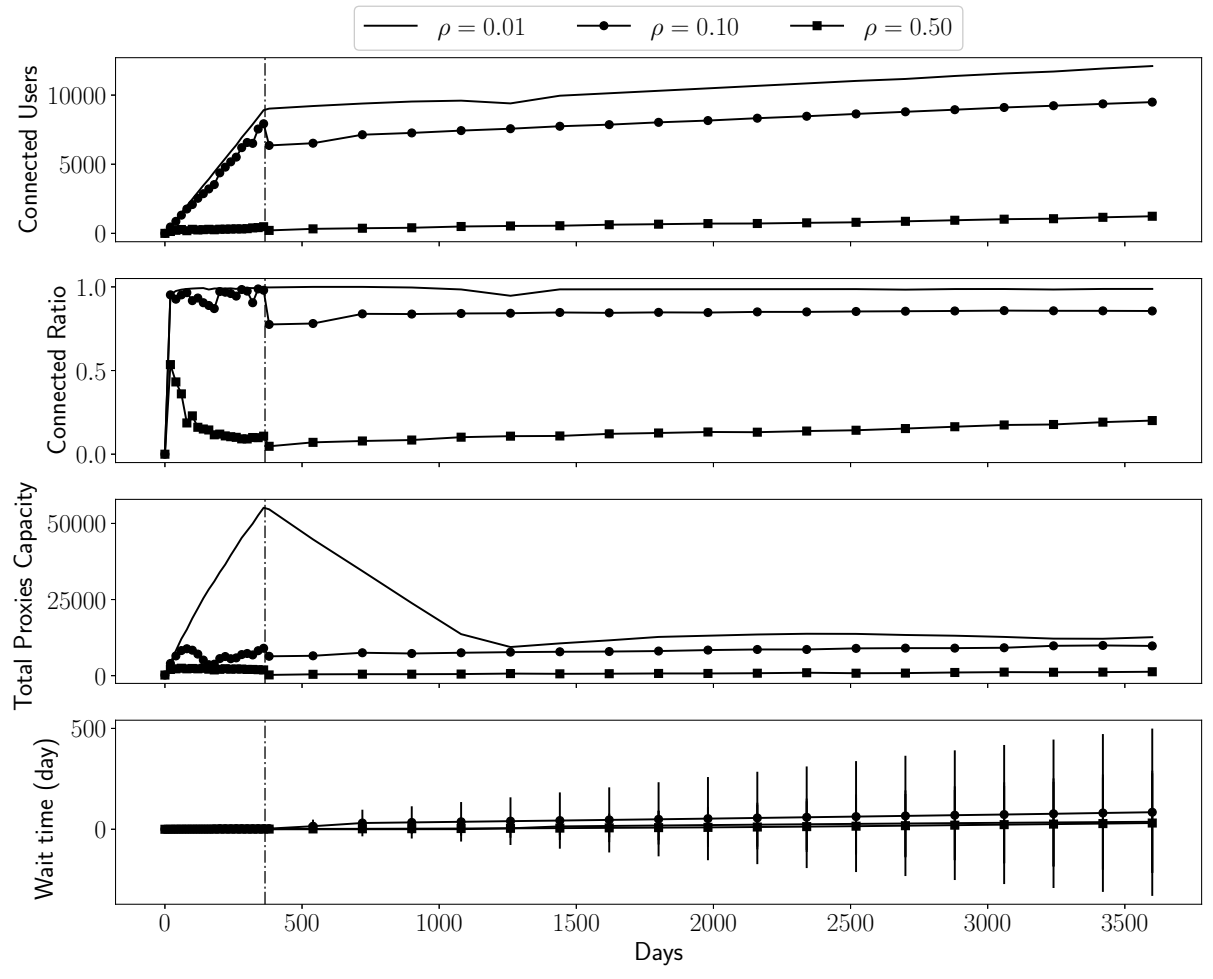

Fig. 7. Omnipresent censor with conservative blocking in Slow world

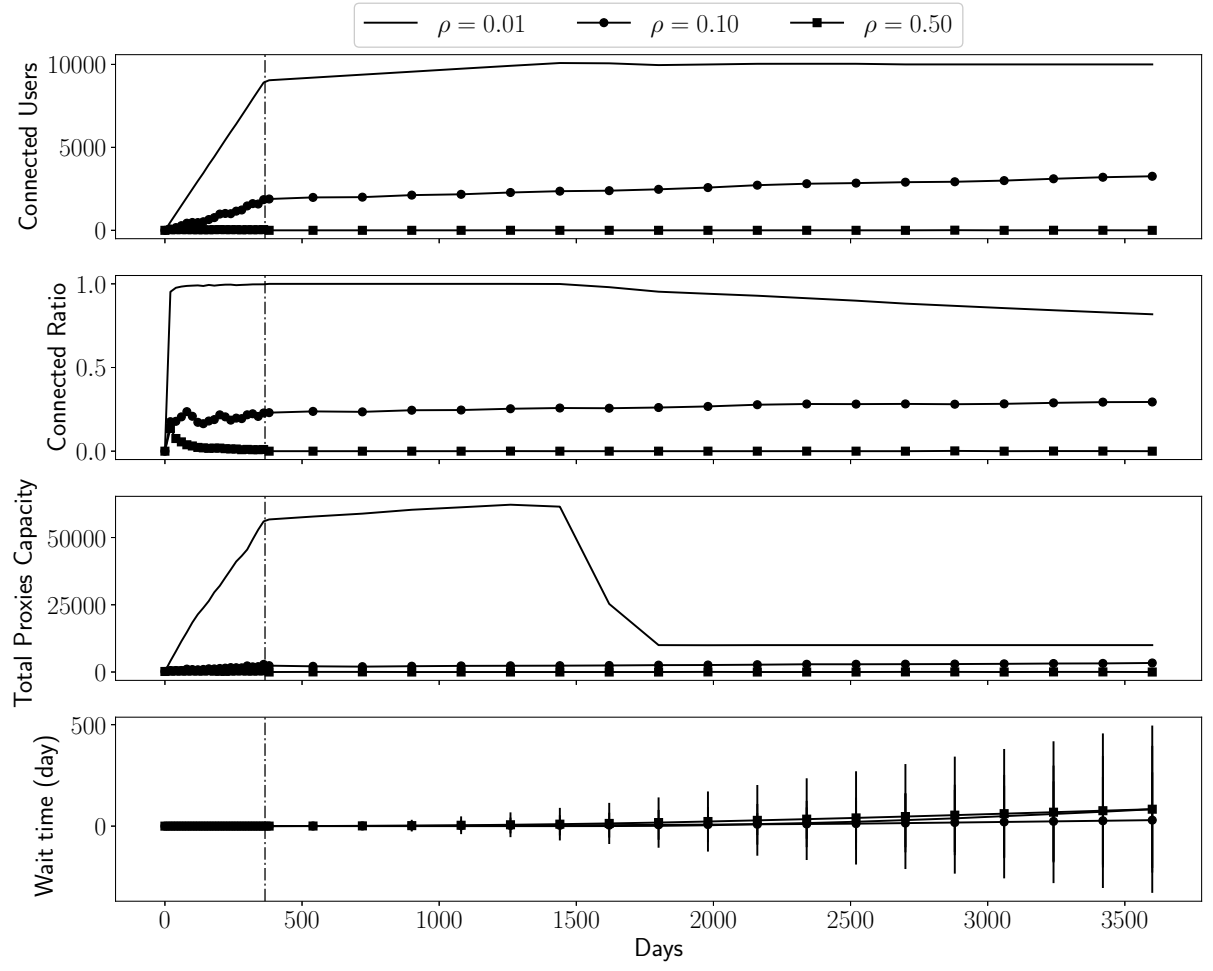

Fig. 8. Omnipresent censor with optimal blocking in Slow world

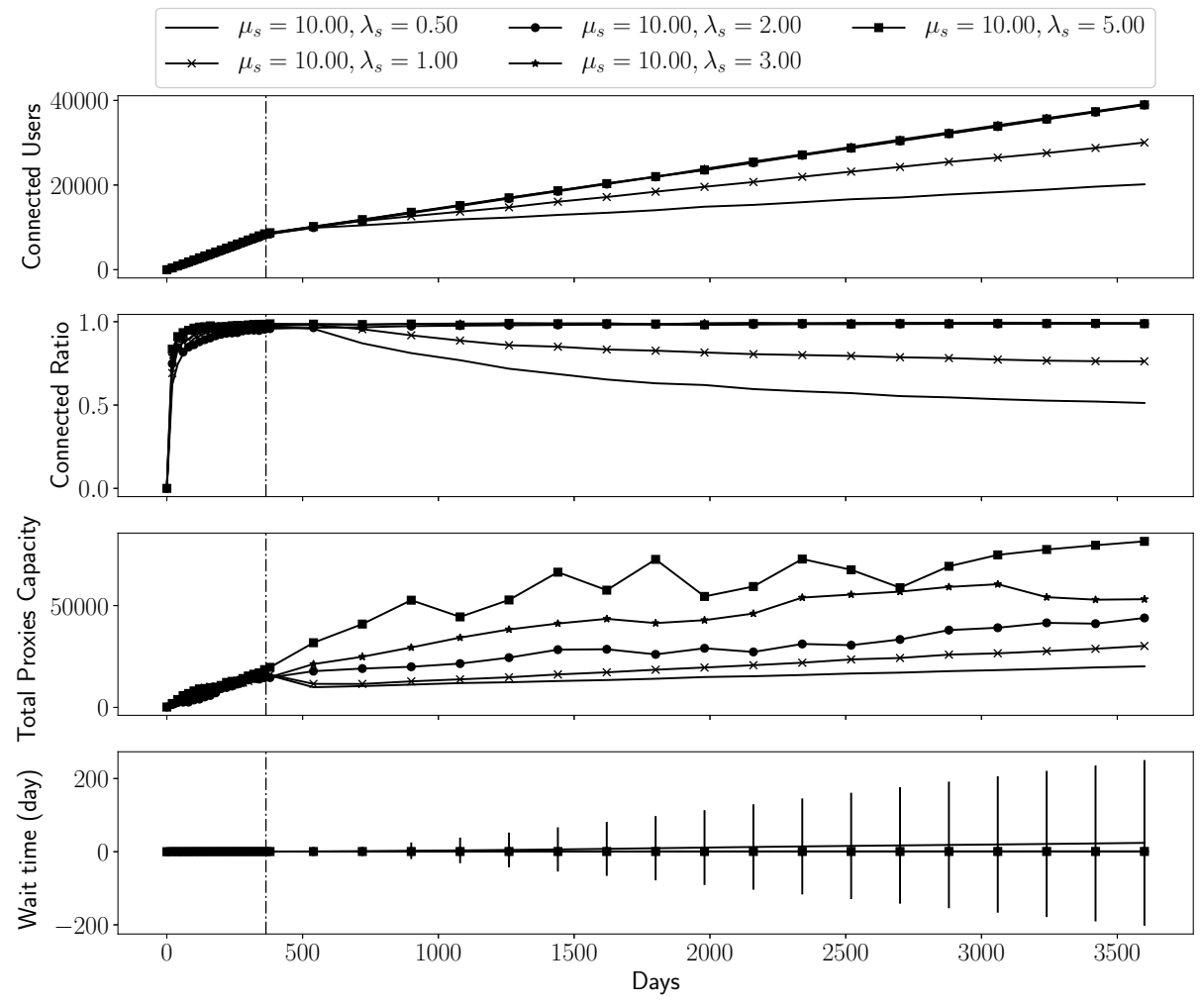

Fig. 9. Omnipresent censor location with optimal blocking in *Alive* world and $\rho = 0.05$

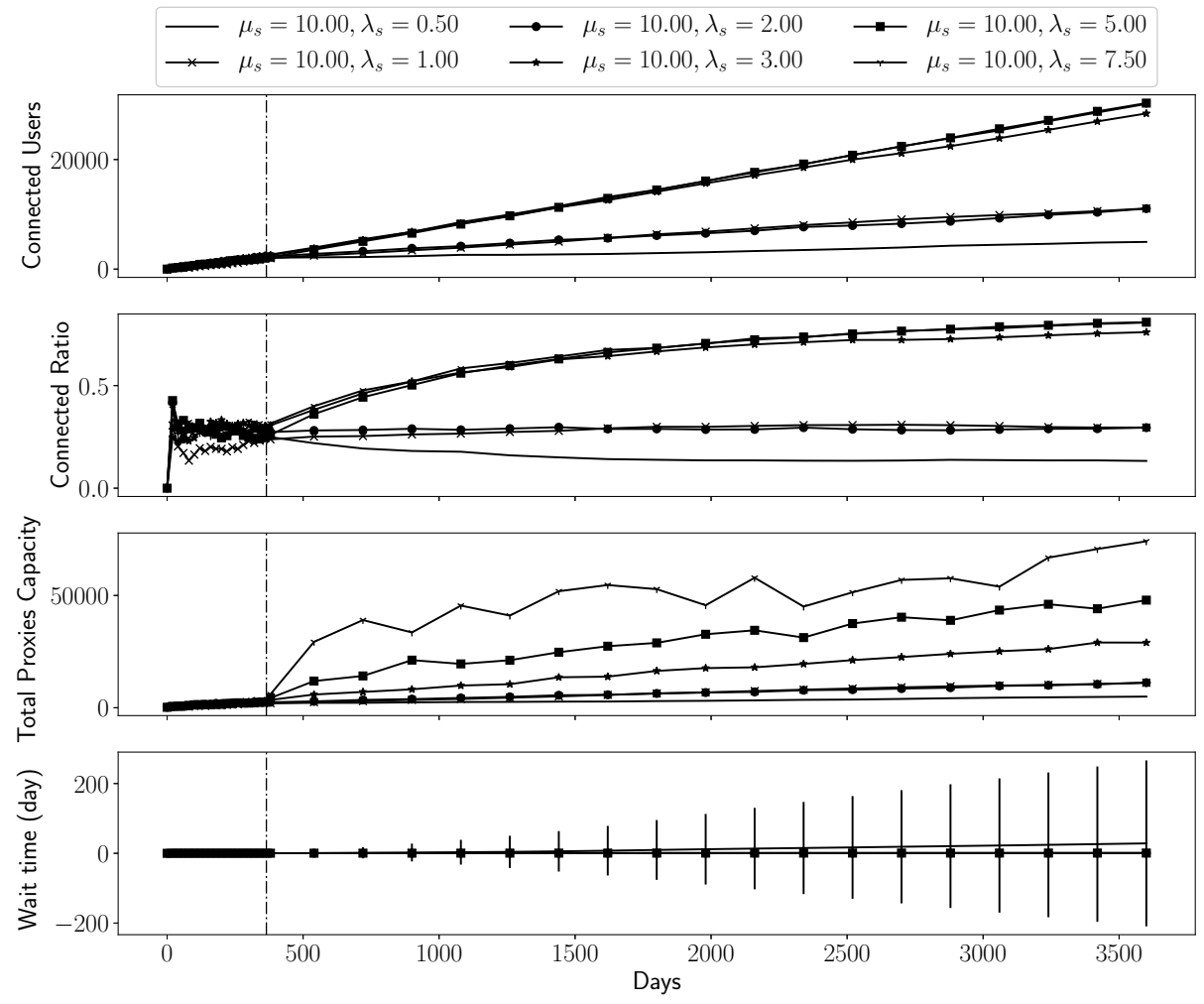

Fig. 10. Omnipresent censor location with optimal blocking in *Alive* world and $\rho = 0.1$

As we can see in Figures 9 and 11, the rate of new proxies, $\lambda_s$, is a critical factor in our system. By comparing different $\lambda_s$ values, we observe that each setting (different $\mu_s$) has an equilibrium point. If $\lambda_s$ is less than that point, the system will degrade over time and the censor will be able to defeat the system eventually, and the speed of degrading is directly related to that rate. On the other hand, if the rate is higher than that equilibrium threshold, the system will be underutilized, and therefore cost-ineffective. Based on these Figures, for high $\lambda_s$, all of the experiments get similar results in the number of connected users, but when $\lambda_s$ is higher than the equilibrium point (e.g., $\lambda_s = 10$) we see a significant unused capacity. Further, in Figures 9 and 11, we show that as we reduce the rate of adding new proxies, the rate of losing proxies increases exponentially.

• **Lesson Four:** The rate of adding new proxies to a circumvention system is crucial for its effectiveness. The rate should be derived based on the capabilities of the censor.

Also, by comparing the Alive (Figures 9 and 10) and Popular (Figures 11 and 12) ecosystems, we see that the birth interval of these systems can have an impact on the long-term operation of the system. If the censoring agents can get into the system at very high rates during the birth interval, the system will not be able to recover, regardless of the rate of adding new proxies during the stable interval. Based on Figures 10 and 12, for a very high rate of new proxies, there is available capacity in the proxies. But the system cannot trust the large number of censored clients who lost many of their proxies during the birth interval. For instance, let us compare the rate $\lambda_s = 10$ to

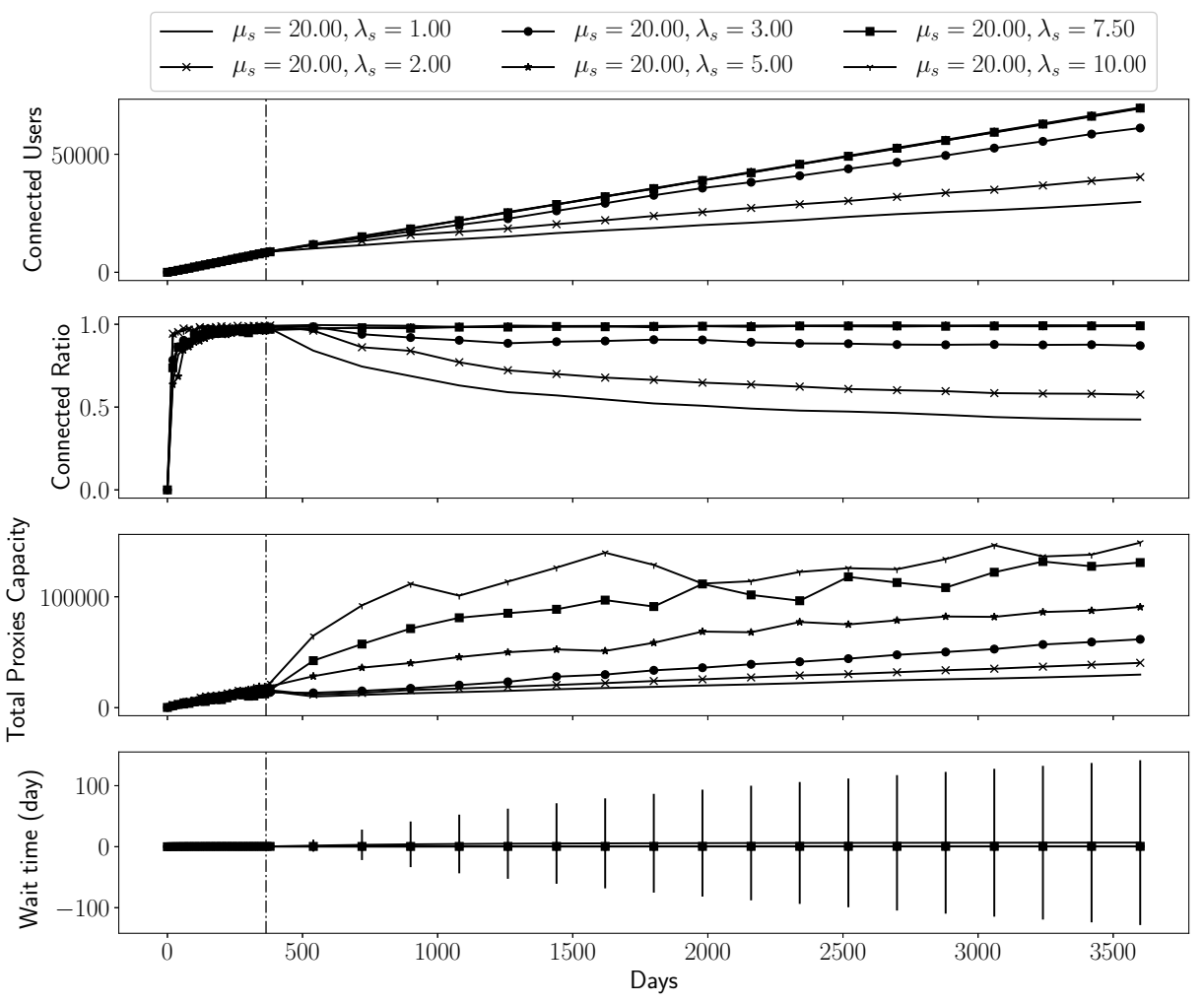

Fig. 11. Omnipresent censor with optimal blocking in *Popular* world and $\rho = 0.05$

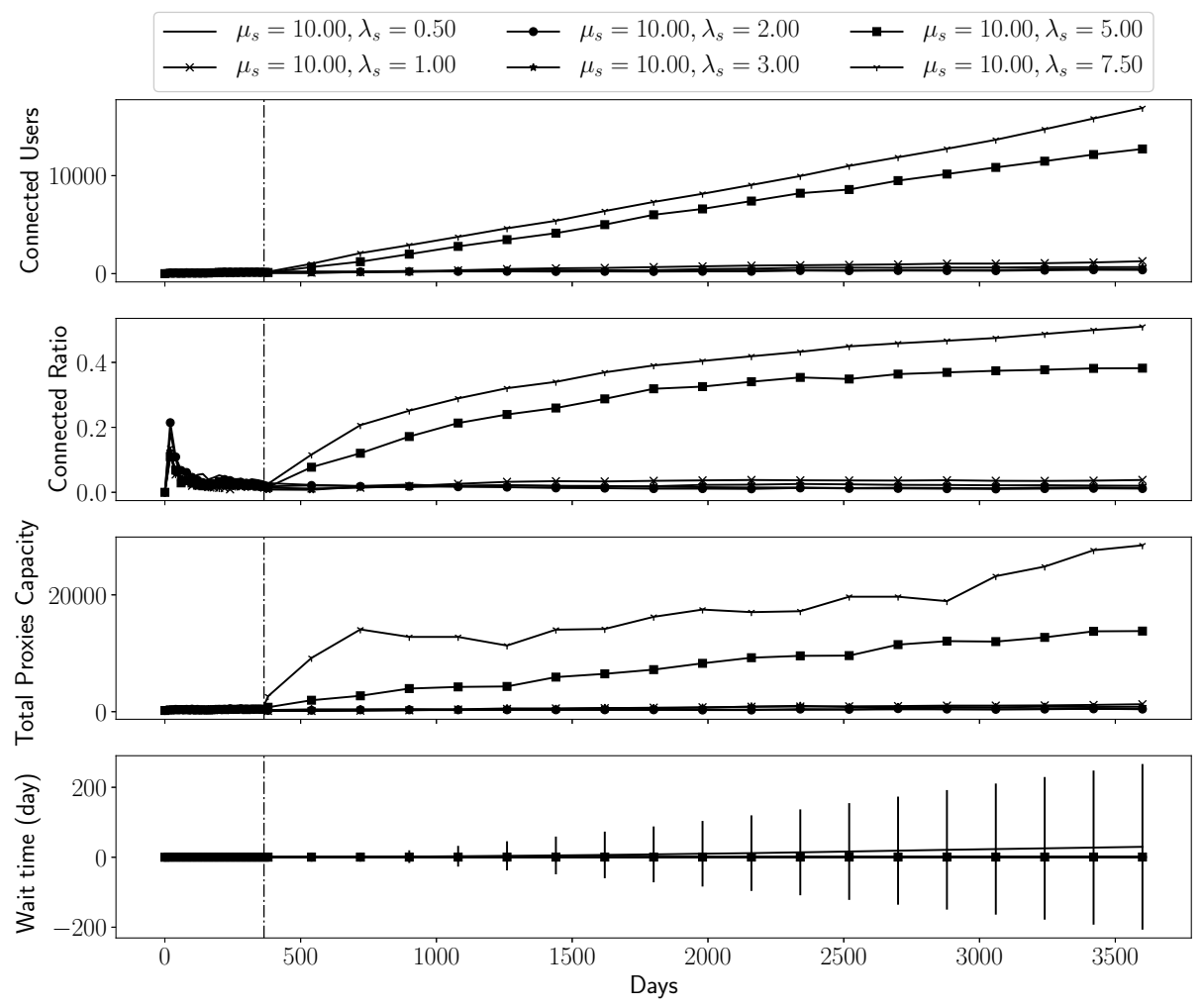

Fig. 13. Omnipresent censor with optimal blocking in *Alive* world and $\rho = 0.2$

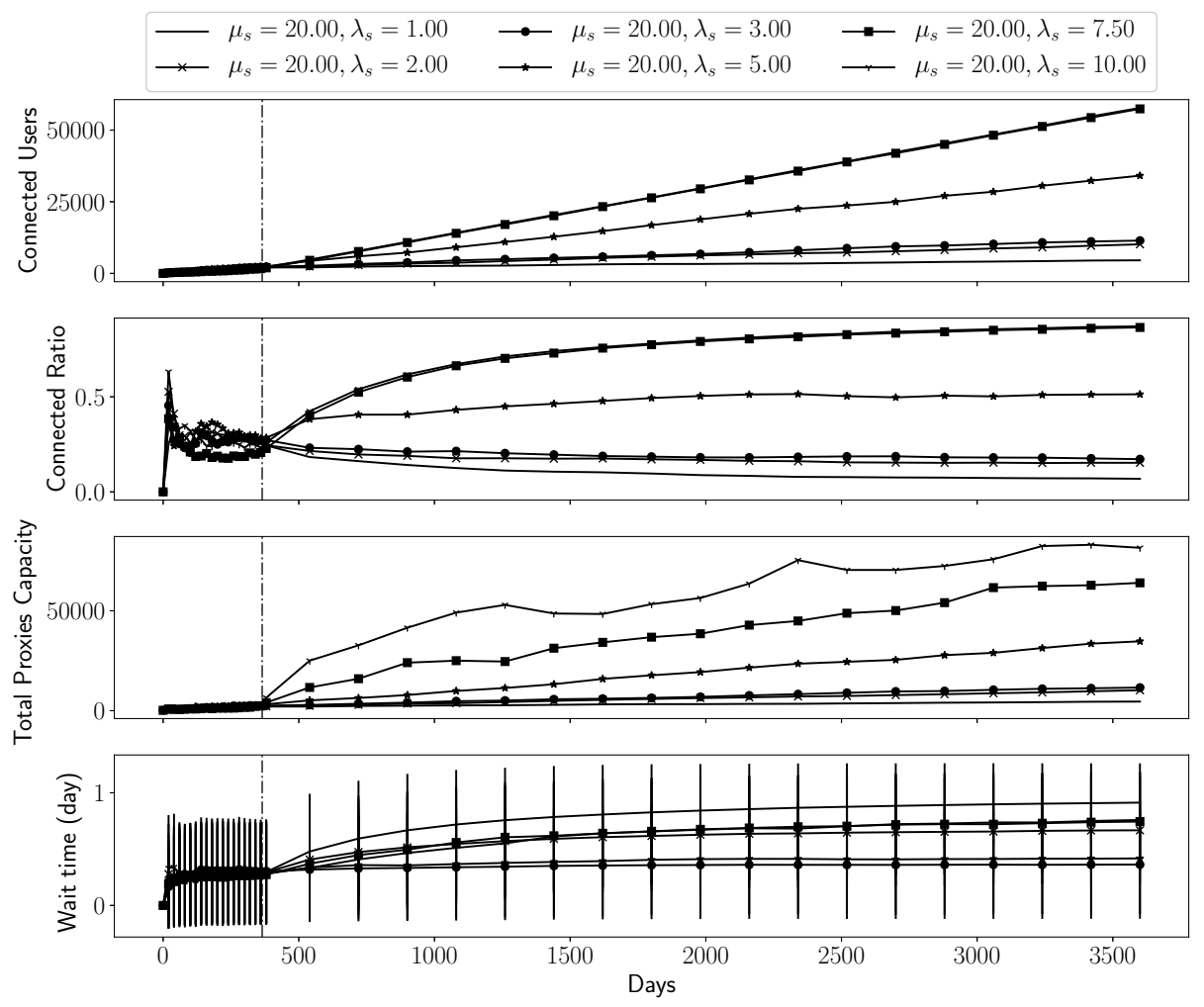

Fig. 12. Omnipresent censor with optimal blocking in *Popular* world and $\rho = 0.1$

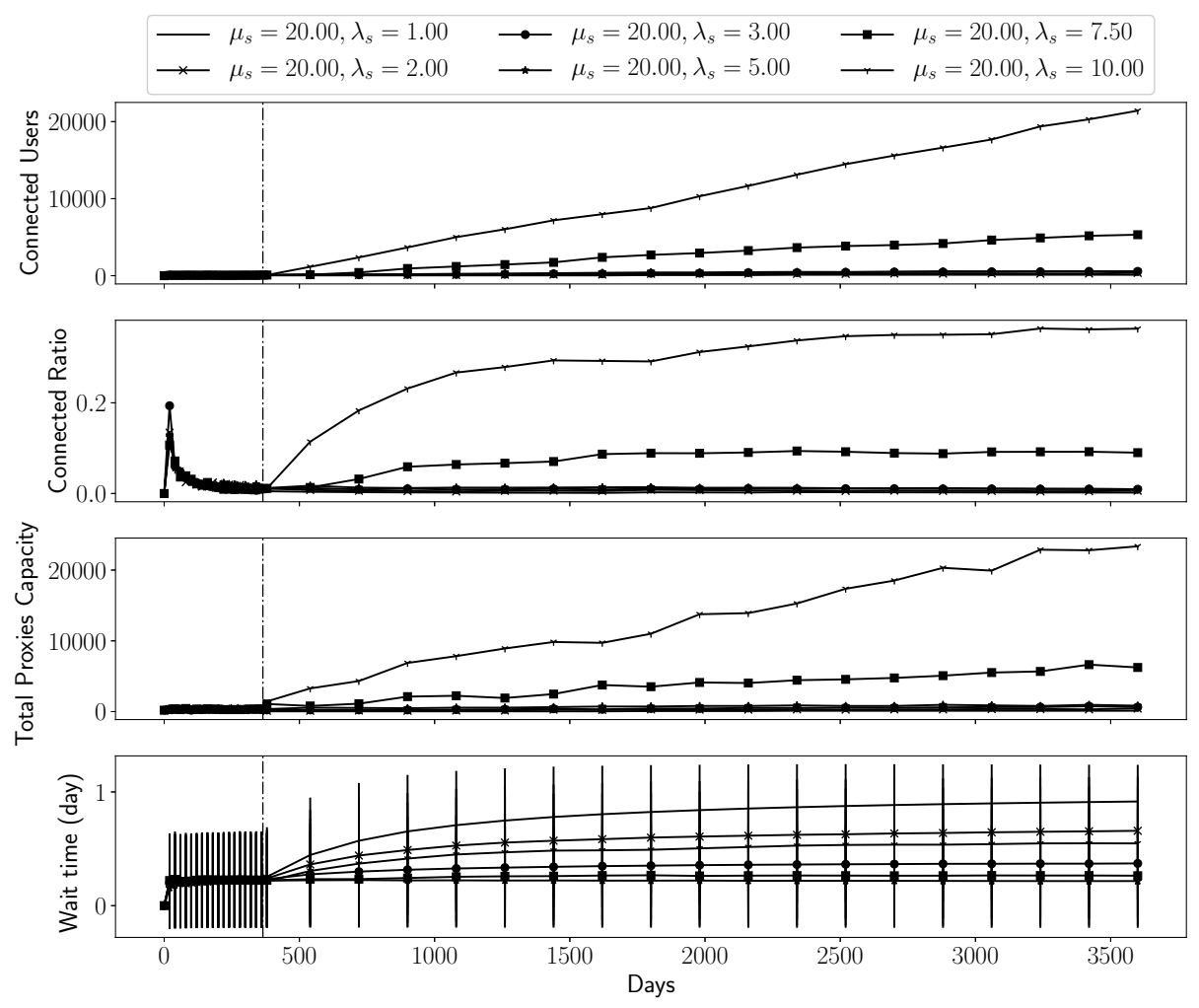

Fig. 14. Omnipresent censor with optimal blocking in *Popular* world and $\rho = 0.2$

$\lambda_s = 7.5$ in Figure 12. In this figure, the total proxy capacity metric is much higher for $\lambda_s = 10$ than $\lambda_s = 7.5$, however, the number of connected users are the same.

• **Lesson Five:** A proxy distribution system should bootstrap with trusted clients. Bootstrapping with a large fraction of malicious (censoring) clients can make the system unrecoverable.

Finally, Figures 13 and 14 compare the Alive and Popular worlds for a high censoring agent rate of $\rho = 0.2$. We see that even for such a high rate of censoring agents, the circumvention system can survive by using a proper rate of adding new proxies to the system.

## *E. Comparison to rBridge*

In Figure 15, we compare our proposed proxy assignment algorithm with rBridge [36], the state of the art prior work. We use an aggressive censor for both of the systems for a fair comparison (rBridge does not have an optimal censor). We choose the parameters similar to values in [36], and we use $\rho = 0.05$, $\mu_s = 5$, $\lambda_s = 0.5$. As can be seen, our proxy distribution mechanism is significantly more resistant to censorship than rBridge (against the same censor attacker). That is, our mechanism can keep a larger number of censored clients unblocked.

• **Lesson Six:** Our game-theoretic proxy distribution mechanism outperforms previous state-of-the-art mechanisms, as

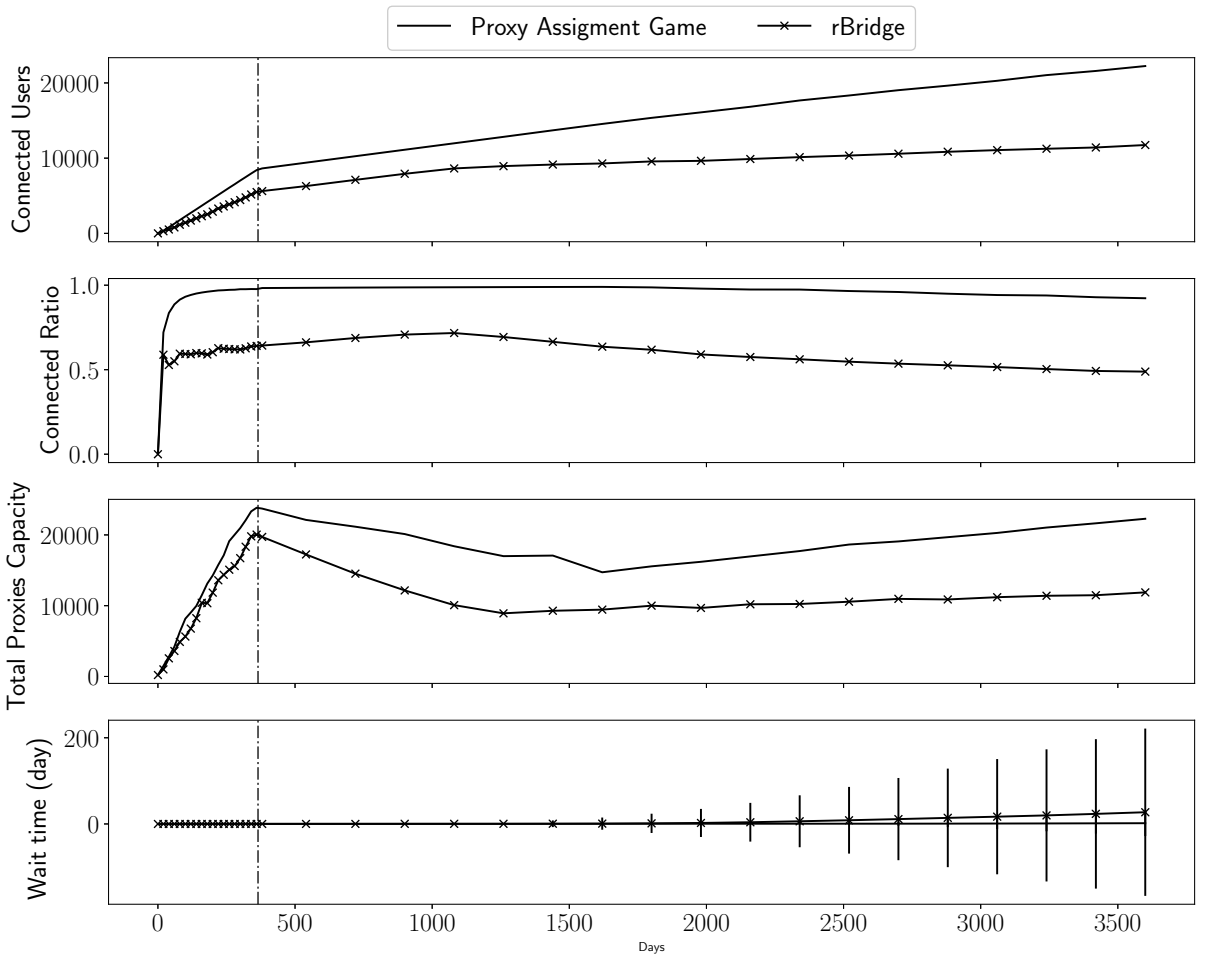

Fig. 15. Comparison with rBridge [36] using the same settings.

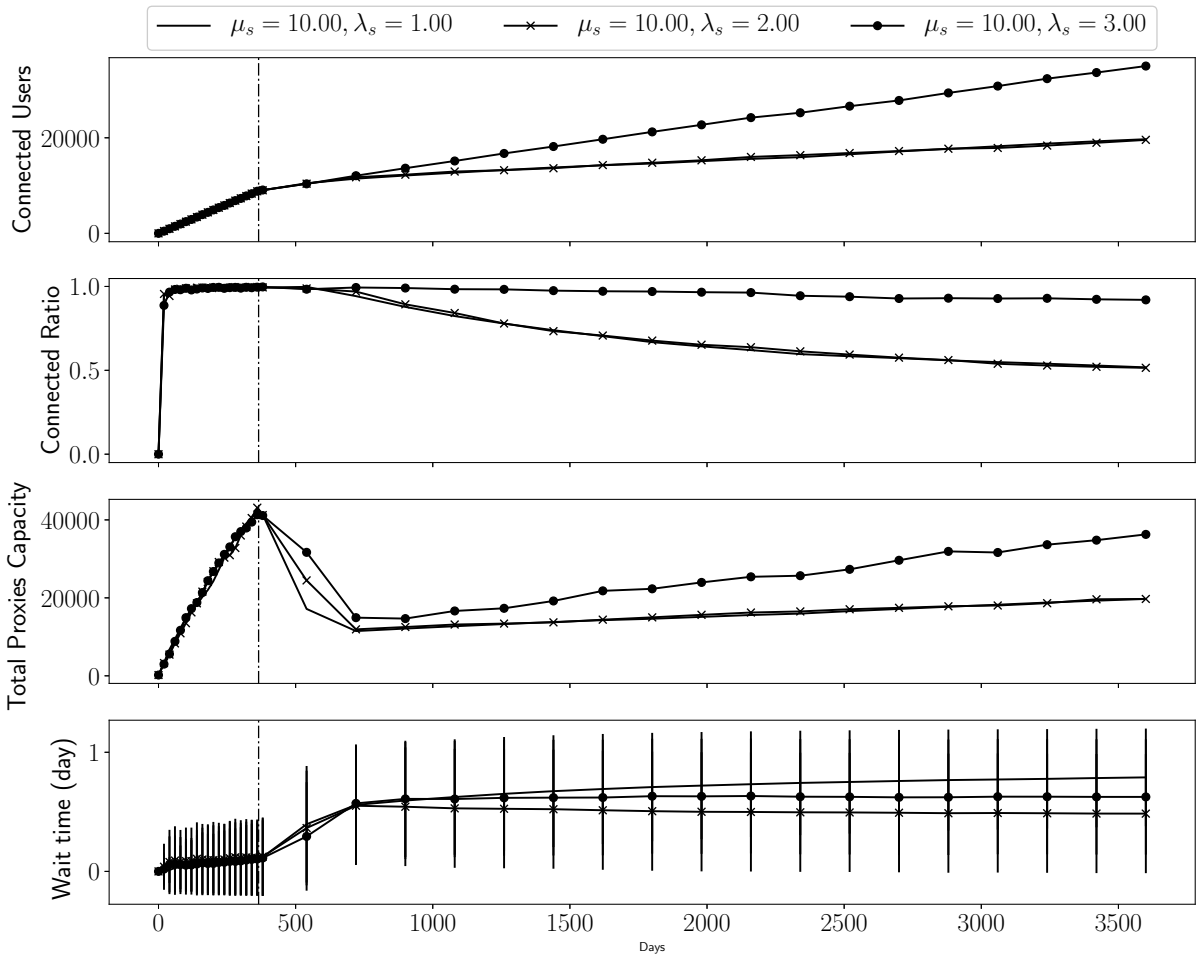

Fig. 16. Omnipresent censor with optimal blocking in *alive* world. We have $\rho = 0.02$ during the birth interval that can be construed as restricted invitation system. In the stable interval, we have $\rho = 0.1$, i.e., open world.

they are based on ad hoc approaches.

## IX. Discussion and Suggestions

One of the main goals of this paper is to analyze the existing proxy distribution mechanisms for censorship circumvention in order to provide insights for designing stronger proxy assignment mechanisms. We have seen in practice that censoring countries are capable of blocking most Tor bridges in their purview [37]. Moreover, the number of Tor bridges has not seen proper increases in recent years (see Figure 2). According to our experiments and analysis, no distribution algorithm can protect the proxies from censoring agents as long as there is no influx of new proxies. Applied to Tor, this means that all bridges can/will eventually be blocked if the rate of new bridges continues to remain low. The reason for the lack of new bridges in the Tor ecosystem can be likely attributed to the fact that adding new bridges to the ecosystem is expensive (for both volunteers and Tor operators). In order to manage this situation and reduce the cost, Tor could change the IP address of bridges, or use expensive technologies like domain fronting [24], [11], [25]. But, nonetheless, Tor as a censorship circumvention tool requires a better policy for bridge distributions. The results of our paper show the importance of central management to play the role of a single distributor in each jurisdiction. Further, our experiments and analysis corroborate the usefulness of our proposed proxy assignment game as a policy for this distributor. Another important factor which is derived from our experiments and can be used in Tor ecosystem is that the rate of new bridges in Tor should be proportional to the ratio of the censoring agents (which is valid for any proxy system).

The other observation from our experiments is the importance of the birth interval. If the censor can corrupt a system during the birth interval, it is very hard for a distributor to recover the system according to our experiments. One of our main suggestions for a proxying system is to use a very restricted invitation system, such as [23], for the birth interval. After a while, i.e., during the stable interval, the system can transit to an open system, without a restricted invitation system. One of the main drawbacks of using an invitation system is that it cannot scale well to a large number of users. But, here, we merely propose using an invitation system in order to control the ecosystem in the birth interval which mitigates the invasion of the censoring agents. Also, most of the invitation systems are capable of handling a fair amount of users in the birth interval. To evaluate our proposal, we designed an additional experiment where the ratios of the censoring agents are different in the birth interval and the stable interval. In the birth interval, we have $\rho = 0.02$. After the birth interval, i.e., stable interval, this ratio increases to $\rho = 0.1$ (for example, by changing to an open registration system). Figure 16 shows the outcome of our experiment. By comparing Figure 16 to Figure 10, we can observe an obvious difference. In other words, the system is able to defend itself during the birth interval, which also means that it is able to maintain its performance afterwards.